\documentclass[10pt,twocolumn,letterpaper]{article}
\usepackage{cvpr}
\usepackage{eso-pic}
\usepackage{times}
\usepackage{graphicx}
\usepackage{amsmath}
\usepackage{amssymb}
\usepackage{cases}
\usepackage{setspace}
\usepackage{overpic}
\usepackage{rotating}
\graphicspath{{figures/}}
\usepackage[sort,numbers]{natbib}
\usepackage{subeqnarray}
\usepackage{multirow}
\usepackage{tabularx}
\usepackage{epstopdf}
\usepackage{subfigure}
\usepackage{enumerate}
\usepackage{here}
\usepackage{bm}
\usepackage{color}
\usepackage{xcolor}
\usepackage{colortbl,booktabs}

\makeatother

\usepackage[pagebackref=true,breaklinks=true,letterpaper=true,colorlinks,bookmarks=false]{hyperref}

\cvprfinalcopy 

\def\cvprPaperID{} 

\begin{document}

\title{Deep Unfolding Network for Image Super-Resolution}

\author{Kai Zhang\qquad\quad  Luc Van Gool\qquad\quad  Radu Timofte\vspace{0.1cm}\\
Computer Vision Lab, ETH Zurich, Switzerland\\
{\tt\small \{kai.zhang, vangool, timofter\}@vision.ee.ethz.ch}\\
\url{https://github.com/cszn/USRNet}
}

\maketitle

\begin{abstract}

Learning-based single image super-resolution (SISR) methods are continuously showing superior effectiveness and efficiency over traditional model-based methods, largely due to the end-to-end training. However, different from model-based methods that can handle the SISR problem with different scale factors, blur kernels and noise levels under a unified MAP (maximum a posteriori) framework, learning-based methods generally lack such flexibility. To address this issue, this paper proposes an end-to-end trainable unfolding network which leverages both learning-based methods and model-based methods. Specifically, by unfolding the MAP inference via a half-quadratic splitting algorithm, a fixed number of iterations consisting of alternately solving a data subproblem and a prior subproblem can be obtained. The two subproblems then can be solved with neural modules, resulting in an end-to-end trainable, iterative network. As a result, the proposed network inherits the flexibility of model-based methods to super-resolve blurry, noisy images for different scale factors via a single model, while maintaining the advantages of learning-based methods. Extensive experiments demonstrate the superiority of the proposed deep unfolding network in terms of flexibility, effectiveness and also generalizability.

\end{abstract}

\section{Introduction}
\label{sec:introduction}

Single image super-resolution (SISR) refers to the process of recovering the natural and sharp detailed high-resolution (HR) counterpart from a low-resolution (LR) image. It is one of the classical ill-posed inverse problems in low-level computer vision and has a wide range of real-world applications, such as enhancing the image visual quality on high-definition displays~\cite{siu2012review,peleg2014statistical} and improving the performance of other high-level vision tasks~\cite{dai2016image}.

Despite decades of studies, SISR still requires further study for academic and industrial purposes~\cite{zhang2015revisiting,li2020group}. The difficulty is mainly caused by the inconsistency between the simplistic degradation assumption of existing SISR methods and the complex degradations of real images~\cite{efrat2013accurate}.
Actually, for a scale factor of $\bf{s}$, the classical (traditional) degradation model of SISR~\cite{elad1997restoration,farsiu2004advances,liu2013bayesian} assumes the LR image $\mathbf{y}$ is a blurred, decimated, and noisy version of an HR image $\mathbf{x}$. Mathematically, it can be expressed by
\begin{equation}\label{eq:sisr_degradation}
  \mathbf{y}\!= \!(\mathbf{x}\otimes \mathbf{k})\!\downarrow_{\bf{s}}\! + \mathbf{n},
\end{equation}
where $\otimes$ represents two-dimensional convolution of $\mathbf{x}$ with blur kernel $\mathbf{k}$, $\downarrow_{\bf{s}}$ denotes the standard $\bf{s}$-fold downsampler, \ie, keeping the upper-left pixel for each distinct $\bf{s}\times\bf{s}$ patch and discarding the others, and $\mathbf{n}$ is usually
assumed to be additive, white Gaussian noise (AWGN) specified by standard deviation (or noise level) $\sigma$~\cite{zhao2016fast}.
With a clear physical meaning, Eq.~\eqref{eq:sisr_degradation} can approximate a variety of LR images by setting proper blur kernels, scale factors and noises for an underlying HR images.
In particular, Eq.~\eqref{eq:sisr_degradation} has been extensively studied in model-based methods which solve a combination of a data term and a prior term under the MAP framework.

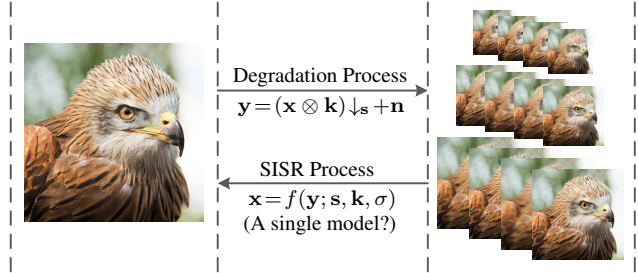
\begin{figure}[!tbp]
\begin{center}
\begin{overpic}[width=0.48\textwidth]{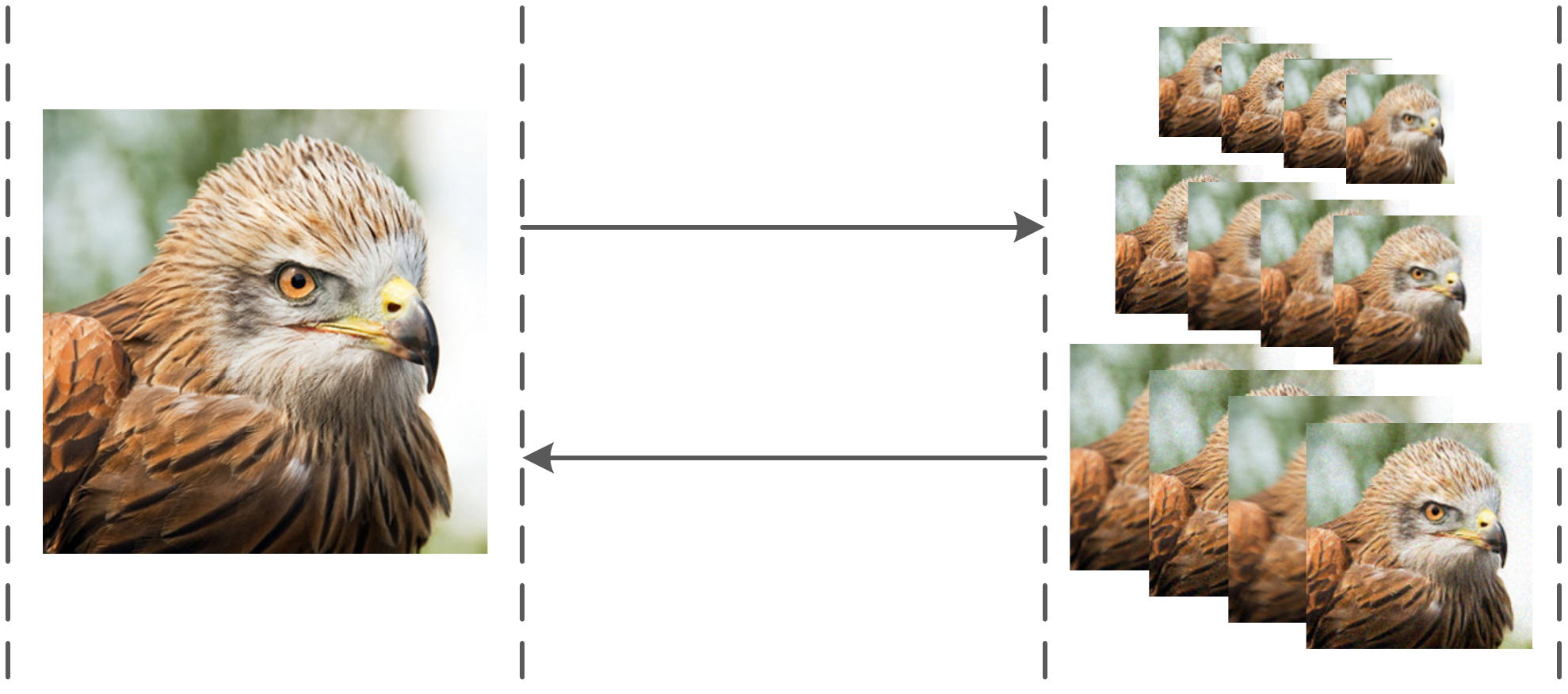}
\put(35.8,31.1){\color{black}{\footnotesize Degradation Process}}
\put(36.5,26.4){\color{black}{\footnotesize $\mathbf{y}\!=\!(\mathbf{x}\otimes\mathbf{k})\!\downarrow_{\bf{s}}\!+\mathbf{n}$}}
\put(40,16.2){\color{black}{\footnotesize SISR Process}}
\put(38.2,11.6){\color{black}{\footnotesize $\mathbf{x}\!=\!f(\mathbf{y};\bf{s},\mathbf{k},\sigma)$}}
\put(37.4,7.7){\color{black}{\footnotesize (A single model?)}}
\end{overpic}
\caption{While a single degradation model (\ie, Eq.~\eqref{eq:sisr_degradation}) can result in various LR images for an HR image, with different blur kernels, scale factors and noise, the study of learning a single deep model to invert all such LR images to HR image is still lacking.}\label{fig:degradation}
\end{center}\vspace{-0.4cm}
\end{figure}

Though model-based methods are usually algorithmically interpretable, they typically lack a standard criterion for their evaluation because, apart from the scale factor, Eq.~\eqref{eq:sisr_degradation} additionally involves a blur kernel and added noise. For convenience, researchers resort to bicubic degradation without consideration of blur kernel and noise level~\cite{yang2008image,timofte2014a+,Dong2016}. However, bicubic degradation is mathematically complicated~\cite{keys1981cubic}, which in turn hinders the development of model-based methods. For this reason, recently proposed SISR solutions are dominated by learning-based methods that learn a mapping function from a bicubicly downsampled LR image to its HR estimation.
Indeed, significant progress on improving PSNR~\cite{kim2015accurate,zhang2018image} and perceptual quality~\cite{ledig2016photo,sajjadi2017enhancenet,wang2018esrgan} for the bicubic degradation has been achieved by learning-based methods, among which  convolutional neural network (CNN) based methods are the most popular, due to their powerful learning capacity and the speed of parallel computing.
Nevertheless, little work has been done on applying CNNs to tackle  Eq.~\eqref{eq:sisr_degradation} via a single model.
Unlike model-based methods, CNNs usually lack flexibility to super-resolve blurry, noisy LR images for different scale factors via a single end-to-end trained model (see Fig.~\ref{fig:degradation}).

In this paper, we propose a deep unfolding super-resolution network (USRNet) to bridge the gap between learning-based methods and model-based methods. On one hand, similar to model-based methods, USRNet can effectively handle the classical degradation model (\ie, Eq.~\eqref{eq:sisr_degradation}) with different blur kernels, scale factors and noise levels via a single model.
On the other hand, similar to learning-based methods, USRNet can be trained in an end-to-end fashion to guarantee effectiveness and efficiency.
To achieve this, we first unfold the model-based energy function via a half-quadratic splitting algorithm. Correspondingly, we can obtain an inference which iteratively alternates between solving two subproblems, one related to a data term and the other to a prior term.
We then treat the inference as a deep network, by replacing the solutions to the two subproblems with neural modules. Since the two subproblems correspond respectively to enforcing degradation consistency knowledge and guaranteeing denoiser prior knowledge, USRNet is well-principled with explicit degradation and prior constraints, which is a distinctive advantage over existing learning-based SISR methods. It is worth noting that since USRNet involves a hyper-parameter for each subproblem, the network contains an additional module for hyper-parameter generation. Moreover, in order to reduce the number of parameters, all the prior modules share the same architecture and same parameters.

The main contributions of this work are as follows:
\begin{itemize}
  \vspace{-0.1cm}
  \item[1)] An end-to-end trainable unfolding super-resolution network (USRNet) is proposed. USRNet is the first attempt to handle the classical degradation model with different scale factors, blur kernels and noise levels via a single end-to-end trained model.
  \vspace{-0.1cm}
  \item[2)] USRNet integrates the flexibility of model-based methods and the advantages of learning-based methods, providing an avenue to bridge the gap between model-based and learning-based methods.
  \vspace{-0.1cm}
  \item[3)] USRNet intrinsically imposes a degradation constraint (\ie, the estimated HR image should accord with the degradation process) and a prior constraint (\ie, the estimated HR image should have natural characteristics) on the solution.
  \vspace{-0.1cm}
  \item[4)] USRNet performs favorably on LR images with different degradation settings, showing great potential for practical applications.
\end{itemize}

\section{Related work}
\label{sec:related_work}

\subsection{Degradation models}
\label{ssc:degradation_models}

Knowledge of the degradation model is crucial for the success of SISR~\cite{yang2014single,efrat2013accurate} because it defines how the LR image is degraded from an HR image. Apart from the classical degradation model and bicubic degradation model, several others have also been proposed in the SISR literature.

In some early works, the degradation model assumes the LR image is directly downsampled from the HR image without blurring, which corresponds to the problem of image interpolation~\cite{caselles1998axiomatic}.
In~\cite{singh2014super,li2017iterative}, the bicubicly downsampled image is further assumed to be corrupted by Gaussian noise or JPEG compression noise. In~\cite{dong2013nonlocally,peleg2014statistical}, the degradation model focuses on Gaussian blurring and a subsequent downsampling with scale factor 3.
Note that, different from Eq.~\eqref{eq:sisr_degradation}, their downsampling keeps the center rather than upper-left pixel for each distinct 3$\times$3 patch.
In~\cite{zhang2018learning}, the degradation model assumes the LR image is the blurred, bicubicly downsampled HR image with some Gaussian noise. By assuming the bicubicly downsampled clean HR image is also clean, \cite{zhang2019deep} treats the degradation model as a composition of deblurring on the LR image and SISR with bicubic degradation.

While many degradation models have been proposed, CNN-based SISR for the classical degradation model has received little attention and deserves further study.

\subsection{Flexible SISR methods}
\label{ssc:flexible_SISR}

Although CNN-based SISR methods have achieved impressive success to handle bicubic degradation, applying them to deal with other more practical degradation models is not straightforward.
For the sake of practicability, it is preferable to design a flexible super-resolver that takes the three key factors, \ie, scale factor, blur kernel and noise level, into consideration.

Several methods have been proposed to tackle bicubic degradation with different scale factors via a single model, such as LapSR~\cite{lai2017deep} with progressive upsampling, MDSR~\cite{lim2017enhanced} with scales-specific branches, Meta-SR~\cite{hu2019meta} with meta-upscale module. To flexibly deal with a blurry LR image, the methods proposed in~\cite{riegler2015conditioned,zhang2018learning} take the PCA dimension reduced blur kernel as input. However, these methods are limited to Gaussian blur kernels.
Perhaps the most flexible CNN-based works which can handle various blur kernels, scale factors and noise levels, are the deep plug-and-play methods~\cite{zhang2017learning,zhang2019deep}. The main idea of such methods is to plug the learned CNN prior into the iterative solution under the MAP framework.  Unfortunately, these are essentially model-based methods which suffer from a high computational burden and they involve manually selected hyper-parameters. How to design an end-to-end trainable model so that better results can be achieved with fewer iterations remains uninvestigated.

While learning-based blind image restoration has recently received considerable attention~\cite{shen2018deep,chen2018fsrnet,yasarla2019deblurring,ren2019neural,lugmayr2019unsupervised},
we note that this work focuses on non-blind SISR which assumes the LR image, blur kernel and noise level are known beforehand. In fact, non-blind SISR is still an active research direction. First, the blur kernel and noise level can be estimated, or are known based on other information (\eg, camera setting). Second, users can control the preference of sharpness and smoothness by tuning the blur kernel and noise level. Third, non-blind SISR can be an intermediate step towards solving blind SISR.

\subsection{Deep unfolding image restoration}
\label{ssc:deep_unfolding}

Apart from the deep plug-and-play methods (see, \eg,~\cite{venkatakrishnan2013plug,chan2016plug,heide2016proximal,brifman2019unified}), deep unfolding methods can also integrate model-based methods and learning-based methods. Their main difference is that the latter optimize the parameters in an end-to-end manner by minimizing the loss function over a large training set, and thus generally produce better results even with fewer iterations.
The early deep unfolding methods can be traced back to~\cite{barbu2009training,samuel2009learning,sun2011learning} where a compact MAP inference based on gradient descent algorithm is proposed for image denoising.
Since then, a flurry of deep unfolding methods based on certain optimization algorithms (\eg, half-quadratic splitting~\cite{afonso2010fast}, alternating direction method of multipliers~\cite{boyd2011distributed} and primal-dual~\cite{chambolle2011first,adler2018learned}) have been proposed to solve different image restoration tasks, such as image denoising~\cite{chen2015trainable,lefkimmiatis2016non}, image deblurring~\cite{schmidt2014shrinkage,kruse2017learning}, image compressive sensing~\cite{sun2016deep,zhang2018ista}, and image demosaicking~\cite{kokkinos2018deep}.

Compared to plain learning-based methods, deep unfolding methods are interpretable and can fuse the degradation constraint into the learning model.
However, most of them suffer from one or several of the following drawbacks. (i) The solution of the prior subproblem without using a deep CNN is not powerful enough for good performance. (ii) The data subproblem is not solved by a closed-form solution, which may hinder convergence. (iii) The whole inference is trained via a stage-wise and fine-tuning manner rather than a complete end-to-end manner.
Furthermore, given that there exists no deep unfolding SISR method to handle the classical degradation model, it is of particular interest to propose such a method that overcomes the above mentioned drawbacks.

\section{Method}
\label{sec:method}

\subsection{Degradation model: classical vs. bicubic}
\label{ssc:degradation_model}

\begin{figure}[!htbp]\footnotesize
\hspace{-0.26cm}
\begin{tabular}{c@{\extracolsep{0.13em}}c@{\extracolsep{0.13em}}c}
        \includegraphics[width=0.16\textwidth]{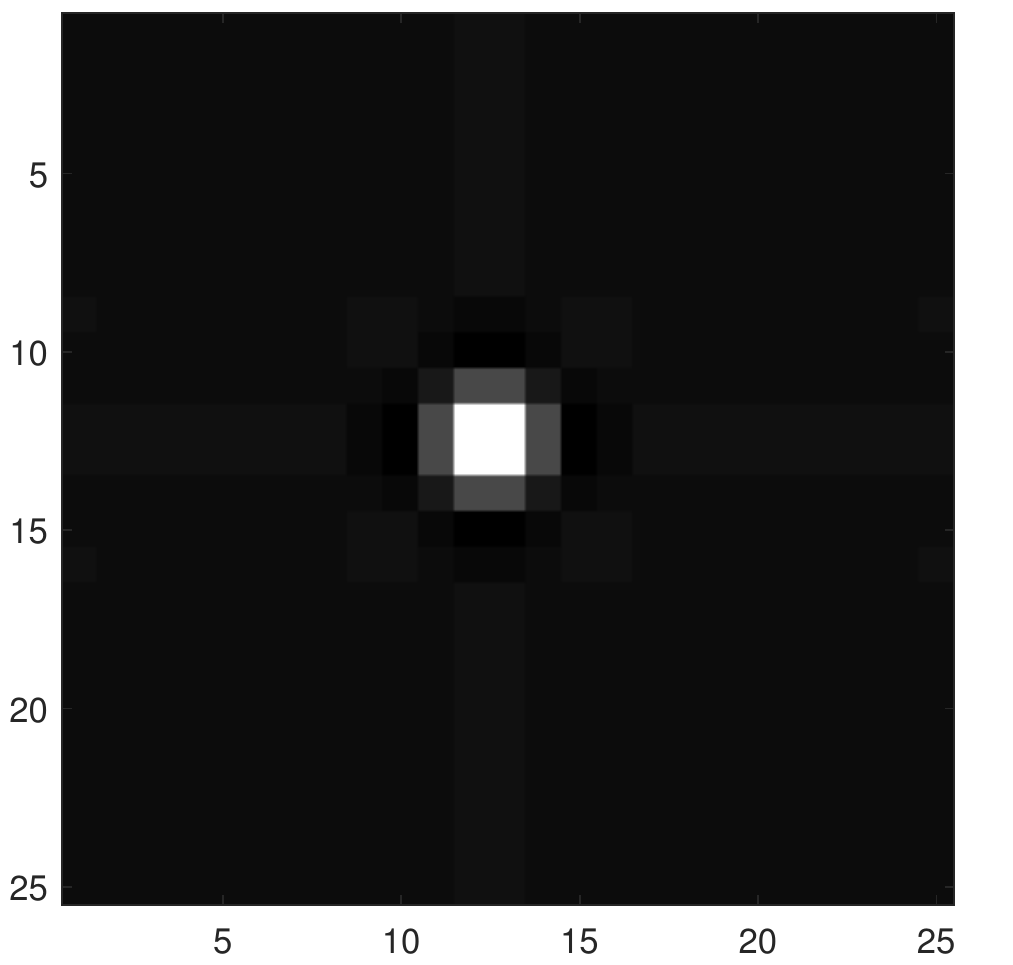}~
		&\includegraphics[width=0.16\textwidth]{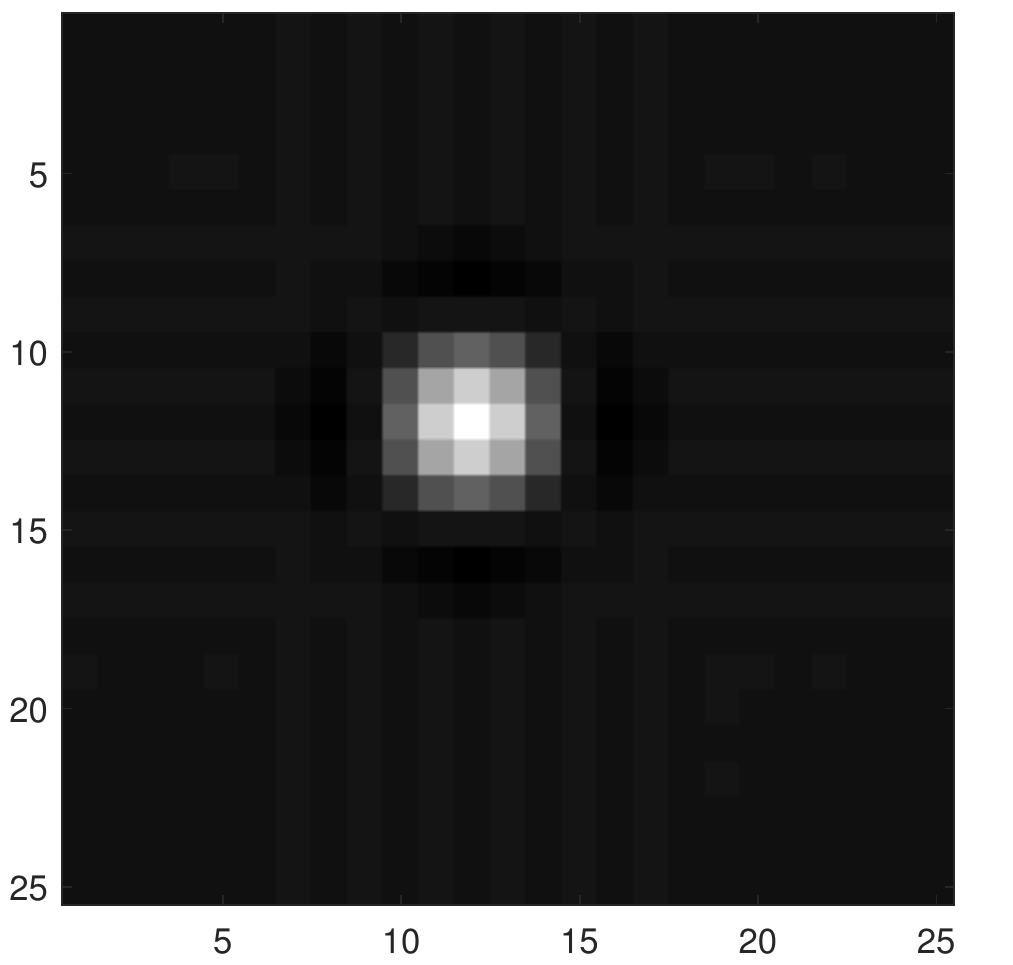}~
		&\includegraphics[width=0.16\textwidth]{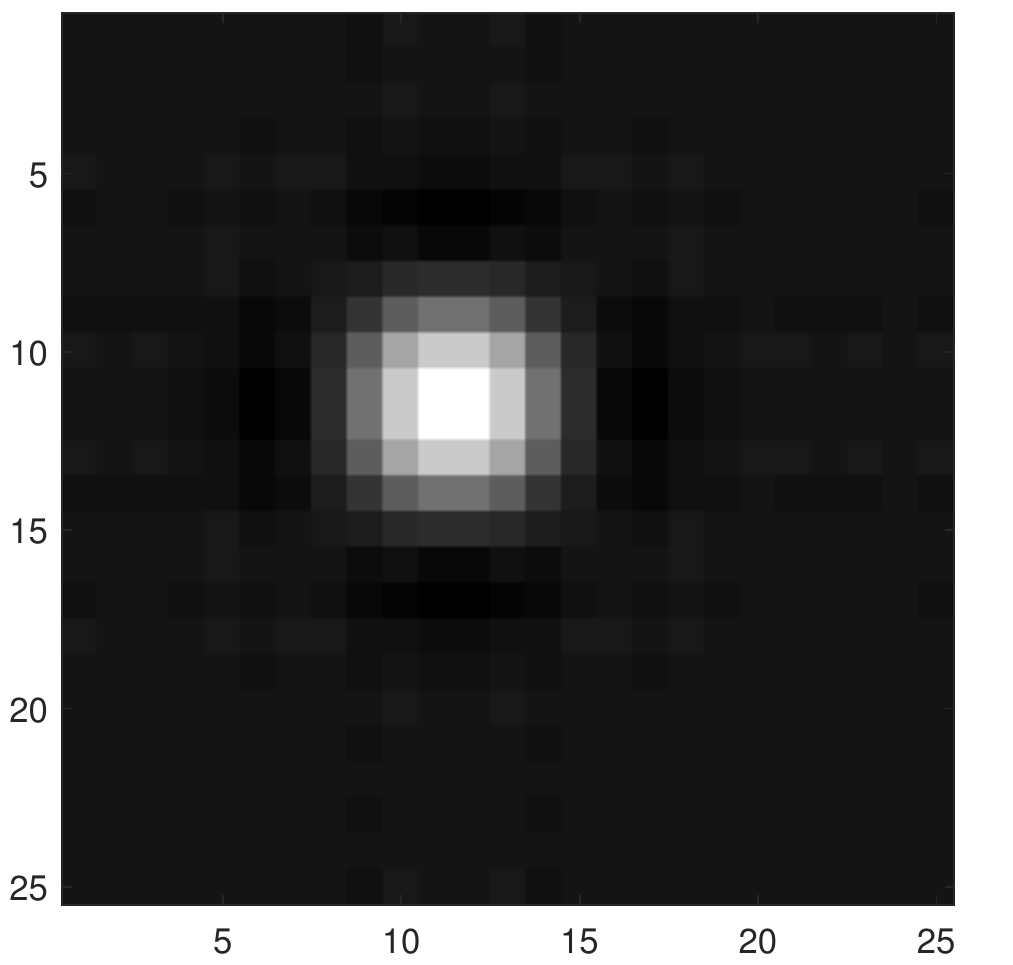}\\
        \includegraphics[width=0.16\textwidth]{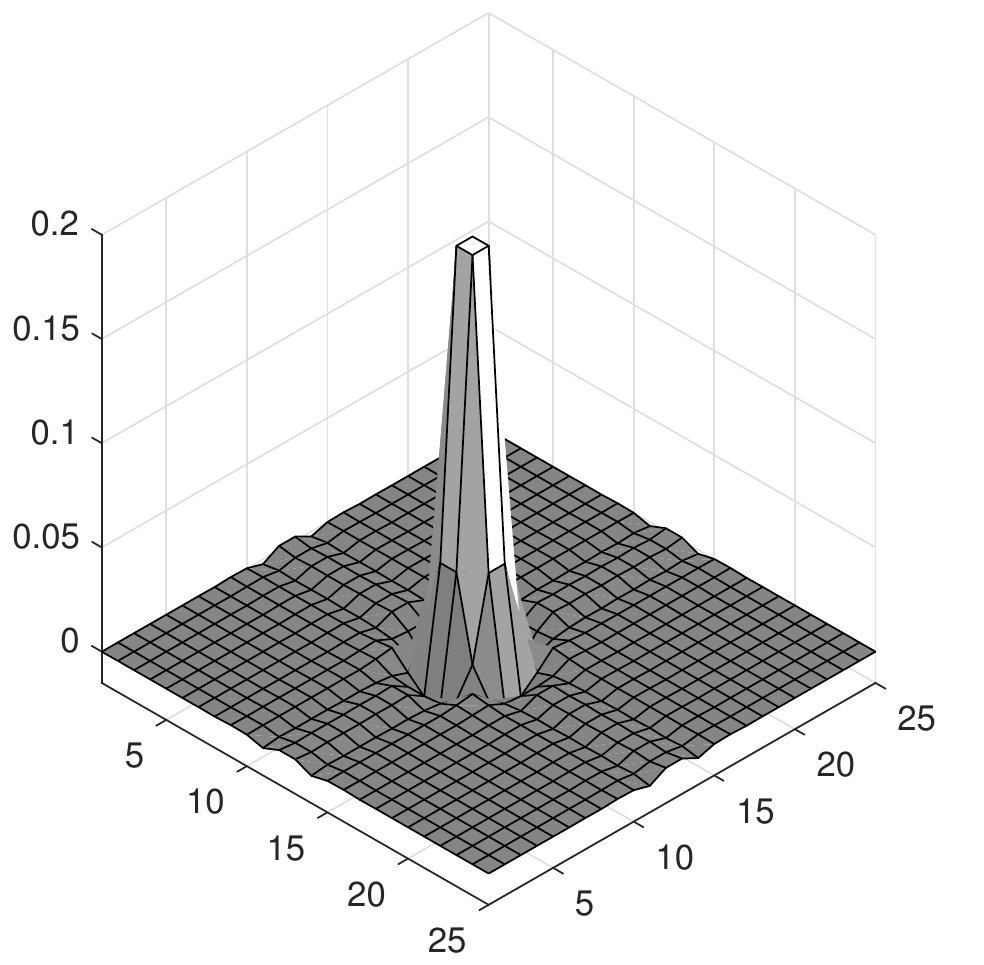}~
		&\includegraphics[width=0.16\textwidth]{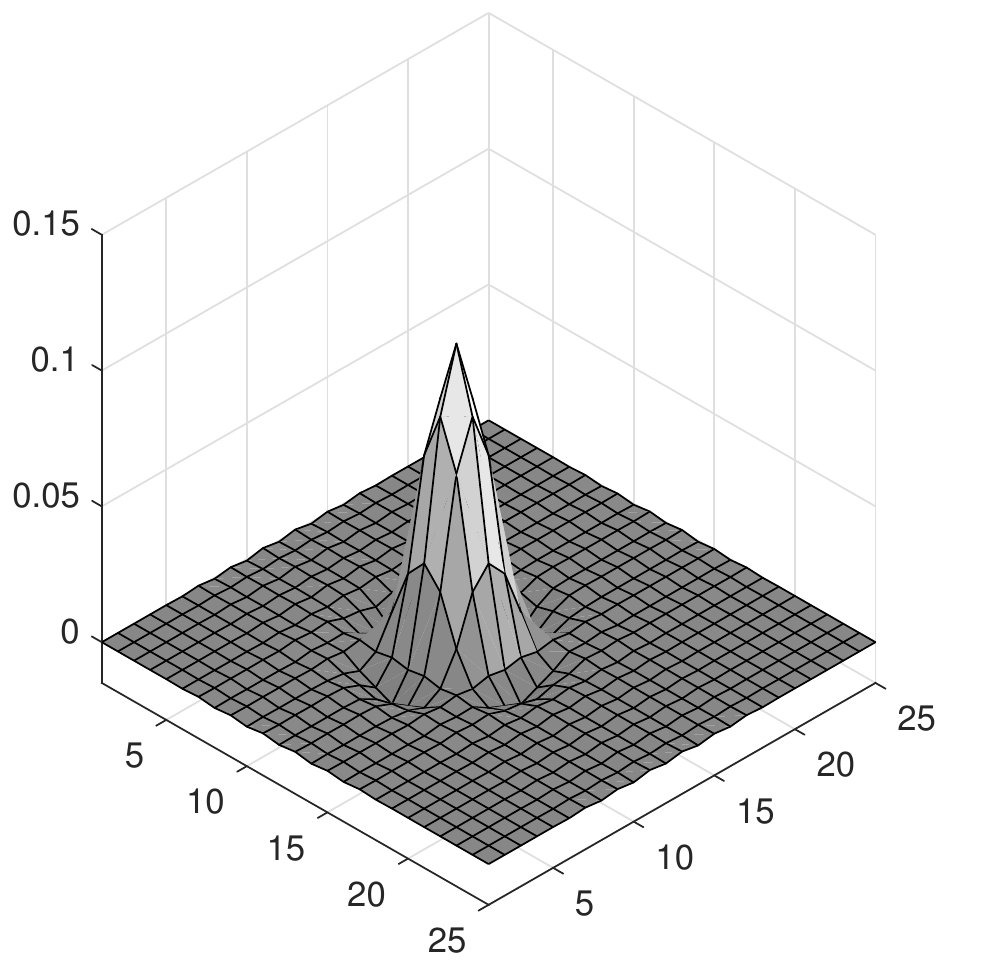}~
		&\includegraphics[width=0.16\textwidth]{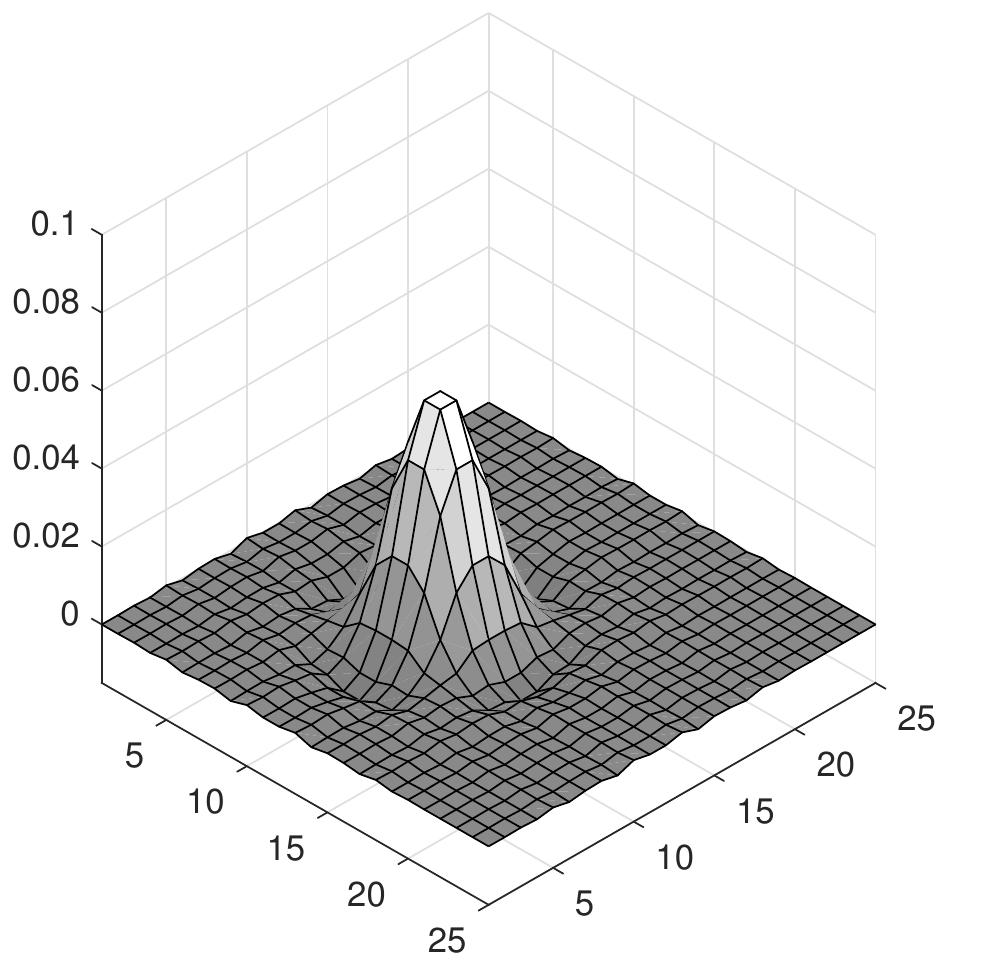}\\

(a) $\mathbf{k}^{\times2}_{bicubic}$ & (b) $\mathbf{k}^{\times3}_{bicubic}$ & (c) $\mathbf{k}^{\times4}_{bicubic}$   \\
	\end{tabular}
    \vspace{0.1cm}
	\caption{Approximated bicubic kernels for scale factors 2, 3 and 4 under the classical SISR degradation model assumption. Note that these kernels contain negative values.}
	\label{fig:bicubic_kernel}
\end{figure}

Since bicubic degradation is well-studied, it is interesting to investigate its relationship to the classical degradation model.
Actually, the bicubic degradation can be approximated by setting a proper blur kernel in Eq.~\eqref{eq:sisr_degradation}.
To achieve this, we adopt the data-driven method to solve the following kernel estimation
problem by minimizing the reconstruction error over a large HR/bicubic-LR pairs $\{(\mathbf{x}$, $\mathbf{y})\}$
\begin{equation}\label{}
  \mathbf{k}^{\times\bf{s}}_{bicubic} = \arg\min\displaystyle_{\mathbf{k}}\|(\mathbf{x}\otimes \mathbf{k})\!\downarrow_{\bf{s}} - \mathbf{y} \|.
\end{equation}
Fig.~\ref{fig:bicubic_kernel} shows the approximated bicubic kernels for scale factors 2, 3 and 4.
It should be noted that since the downsamlping operation selects the upper-left pixel for each distinct $\bf{s}\times\bf{s}$ patch, the bicubic kernels for scale factors 2, 3 and 4 have a center shift of 0.5, 1 and 1.5 pixels to the upper-left direction, respectively.

\subsection{Unfolding optimization}\label{section:deepunfolding}
According to the MAP framework, the HR image could be estimated by minimizing the following energy function
\begin{equation}\label{eq_map}
  E(\mathbf{x}) = \frac{1}{2\sigma^2}\|\mathbf{y} - (\mathbf{x}\otimes\mathbf{k})\!\downarrow_{\bf{s}} \|^2 + \lambda \Phi(\mathbf{x}),
\end{equation}
where $\frac{1}{2\sigma^2}\|\mathbf{y}-(\mathbf{x}\otimes\mathbf{k})\!\!\downarrow_{\bf{s}})\|^2$ is the data term, $\Phi(\mathbf{x})$ is the prior term, and $\lambda$ is a trade-off parameter.
In order to obtain an unfolding inference for Eq.~\eqref{eq_map}, the half-quadratic splitting (HQS) algorithm is selected due to its simplicity and fast convergence in many applications. HQS tackles Eq.~\eqref{eq_map} by introducing an auxiliary variable $\mathbf{z}$, leading to the following approximate equivalence
\begin{equation}\label{eq_mapq}
  E_\mu(\mathbf{x}, \mathbf{z}) = \frac{1}{2\sigma^2}\|\mathbf{y}-(\mathbf{z}\otimes \mathbf{k})\!\downarrow_{\bf{s}}\|^2 + \lambda\Phi(\mathbf{x}) + \frac{\mu}{2}\|\mathbf{z} - \mathbf{x}\|^2,
\end{equation}
where $\mu$ is the penalty parameter. Such problem can be addressed
by iteratively solving subproblems for $\mathbf{x}$ and $\mathbf{z}$
\begin{numcases}{}
\!\!\!\mathbf{z}_{k}\!=\!\arg\min\displaystyle_{\mathbf{z}}\|\mathbf{y}-(\mathbf{z}\otimes \mathbf{k})\!\downarrow_{\bf{s}}\!\|^2 \!+\! \mu\sigma^2\|\mathbf{z}-\mathbf{x}_{k-1}\|^2, \quad \label{eq_s1}\\
\!\!\!\mathbf{x}_{k}\!=\!\arg\min\displaystyle_{\mathbf{x}}\frac{\mu}{2}\|\mathbf{z}_{k}-\mathbf{x}\|^2 + \lambda\Phi(\mathbf{x}). \label{eq_s2}
\end{numcases}
According to Eq.~\eqref{eq_s1}, $\mu$ should be large enough so that $\mathbf{x}$ and $\mathbf{z}$ are approximately equal to the fixed point. However, this would also result in slow convergence.
Therefore, a good rule of thumb is to iteratively increase $\mu$.
For convenience, the $\mu$ in the $k$-th iteration is denoted by $\mu_k$.

It can be observed that the data term and the prior term are decoupled into Eq.~\eqref{eq_s1} and Eq.~\eqref{eq_s2}, respectively.
For the solution of Eq.~\eqref{eq_s1}, the fast Fourier transform (FFT) can be utilized by assuming the convolution is carried out with circular boundary conditions. Notably, it has a closed-form expression~\cite{zhao2016fast}
\begin{equation}\label{eq:sisr}
  \mathbf{z}_{k} \!=\! \mathcal{F}^{-1}\!\left(\!\frac{1}{\alpha_k}\Big(\mathbf{d}\! -\! \overline{\mathcal{F}(\mathbf{k})} \odot_{\bf{s}}\frac{(\mathcal{F}(\mathbf{k})\mathbf{d})\Downarrow_{\bf{s}} }{(\overline{\mathcal{F}(\mathbf{k})}\mathcal{F}(\mathbf{k}))\Downarrow_{\bf{s}} \!+\alpha_k}\Big)\!\right)
\end{equation}
where $\mathbf{d}$ is defined as
$$\mathbf{d} = \overline{\mathcal{F}(\mathbf{k})}\mathcal{F}(\mathbf{y}\uparrow_{\bf{s}}) + \alpha_{k}\mathcal{F}(\mathbf{x}_{k-1})$$
with $\alpha_k\triangleq\mu_k\sigma^2$  and where the $\mathcal{F}(\cdot)$ and $\mathcal{F}^{-1}(\cdot)$ denote FFT and inverse FFT, $\overline{\mathcal{F}(\cdot)}$ denotes complex
conjugate of $\mathcal{F}(\cdot)$, $\odot_{\bf{s}}$ denotes the distinct block processing operator with element-wise multiplication, \ie, applying element-wise multiplication to the $\bf{s}\times\bf{s}$ distinct blocks of $\overline{\mathcal{F}(\mathbf{k})}$, $\Downarrow_{\bf{s}}$ denotes the distinct block downsampler, \ie, averaging the $\bf{s}$$\times$$\bf{s}$ distinct blocks, $\uparrow_{\bf{s}}$ denotes the standard ${\bf{s}}$-fold upsampler, \ie, upsampling the spatial size by filling the new entries with zeros.
It is especially noteworthy that Eq.~\eqref{eq:sisr} also works for the special case of deblurring when ${\bf{s}}=1$.
For the solution of Eq.~\eqref{eq_s2}, it is known that, from a Bayesian perspective, it actually corresponds to a denoising problem with noise level $\beta_k \triangleq \sqrt{\lambda/\mu_k}$~\cite{chan2016plug}.

\subsection{Deep unfolding network}
\label{section:deepnetwork}
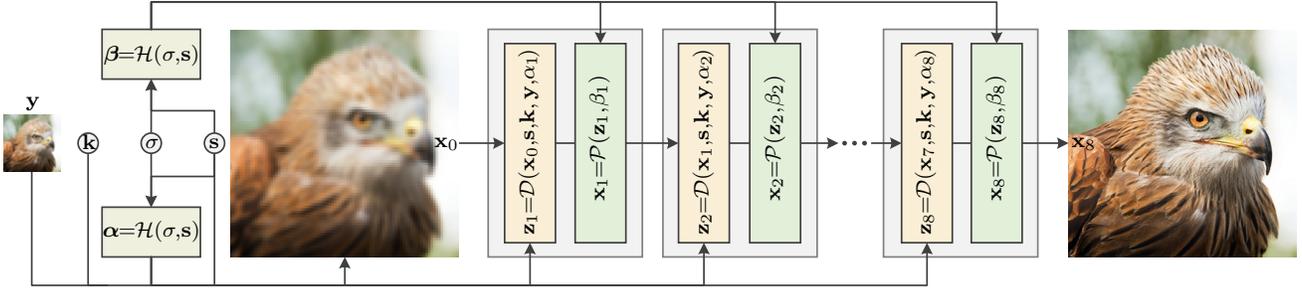
\begin{figure*}[!tbp]
\begin{center}
\begin{overpic}[width=0.99\textwidth]{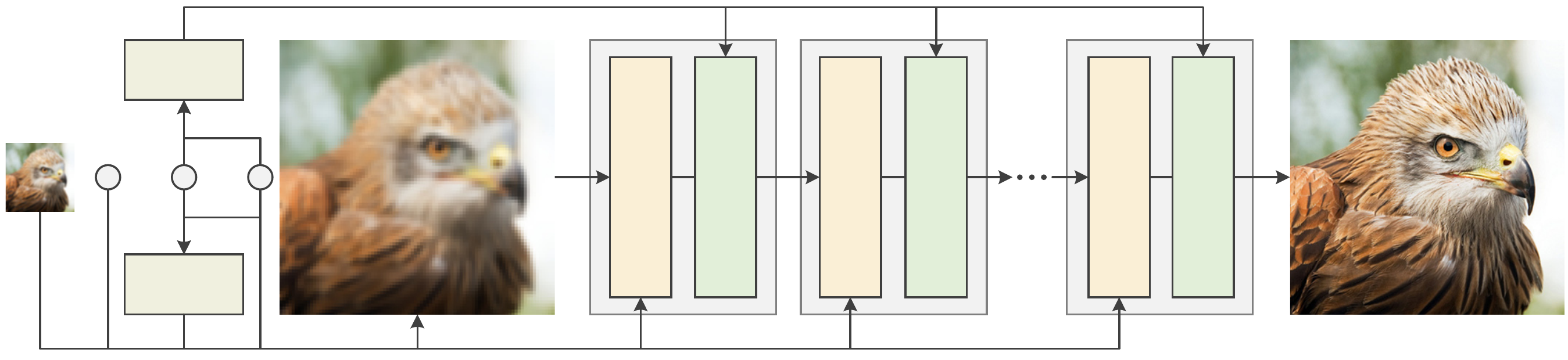}
\put(2.1,14.50){\color{black}{\footnotesize $\mathbf{y}$}}
\put(8.1,18){\color{black}{\footnotesize $\bm{\beta}\!\!=\!\!\mathcal{H}(\sigma, \!\bf{s})$}}
\put(6.45,11.20){\color{black}{\footnotesize $\mathbf{k}$}}
\put(11.3,11.3){\color{black}{\footnotesize $\sigma$}}
\put(16.2,11.3){\color{black}{\footnotesize $\bf{s}$}}
\put(33.5,11.35){\color{black}{\footnotesize $\mathbf{x}_0$}}
\put(8.0,4.5){\color{black}{\footnotesize $\bm{\alpha}\!\!=\!\!\mathcal{H}(\sigma, \!\bf{s})$}}
\put(82.5,11.35){\color{black}{\footnotesize $\mathbf{x}_8$}}
\begin{turn}{90}
\put(4.6,-41.2){\color{black}{\footnotesize $\mathbf{z}_{1}\!\!=\!\!\mathcal{D}(\mathbf{x}_{0}, \!\mathbf{s}, \!\mathbf{k}, \mathbf{y}, \!\alpha_{1}\!)$}}
\put(7.2,-46.6){\color{black}{\footnotesize $\mathbf{x}_{1}\!\!=\!\!\mathcal{P}(\mathbf{z}_{1}, \!\beta_1)\!$}}

\put(4.6,-54.6){\color{black}{\footnotesize $\mathbf{z}_{2}\!\!=\!\!\mathcal{D}(\mathbf{x}_{1}, \!\mathbf{s}, \!\mathbf{k}, \mathbf{y}, \!\alpha_{2}\!)$}}
\put(7.2,-60.1){\color{black}{\footnotesize $\mathbf{x}_{2}\!\!=\!\!\mathcal{P}(\mathbf{z}_{2}, \!\beta_2)\!$}}

\put(4.6,-71.7){\color{black}{\footnotesize $\mathbf{z}_{8}\!\!=\!\!\mathcal{D}(\mathbf{x}_{7}, \!\mathbf{s}, \!\mathbf{k}, \mathbf{y}, \!\alpha_{8}\!)$}}
\put(7.2,-77.1){\color{black}{\footnotesize $\mathbf{x}_{8}\!\!=\!\!\mathcal{P}(\mathbf{z}_{8}, \!\beta_8)\!$}}
\end{turn}
\end{overpic}
\end{center}
\caption{The overall architecture of the proposed USRNet with $K=8$ iterations. USRNet can flexibly handle the classical degradation (\ie, Eq.~\eqref{eq:sisr_degradation}) via a single model as it takes the LR image $\mathbf{y}$, scale factor $\bf{s}$, blur kernel $\mathbf{k}$ and noise level $\sigma$ as input. Specifically, USRNet consists of three main modules, including the data module $\mathcal{D}$ that makes HR estimation clearer, the prior module $\mathcal{P}$ that makes HR estimation cleaner, and the hyper-parameter module $\mathcal{H}$ that controls the outputs of $\mathcal{D}$ and $\mathcal{P}$.}
\label{fig:architecture}
\end{figure*}

Once the unfolding optimization is determined, the next step is to design the unfolding super-resolution network (USRNet). Because the unfolding optimization mainly consists of iteratively solving a data subproblem (\ie, Eq.~\eqref{eq_s1}) and a prior subproblem (\ie, Eq.~\eqref{eq_s2}), USRNet should alternate between a data module $\mathcal{D}$ and a prior module $\mathcal{P}$.
In addition, as the solutions of the subproblems also take the hyper-parameters $\alpha_k$ and $\beta_k$ as input, respectively, a hyper-parameter module $\mathcal{H}$ is further introduced into USRNet. Fig.~\ref{fig:architecture} illustrates the overall architecture of USRNet with $K$ iterations, where $K$ is empirically set to 8 for the speed-accuracy trade-off.
Next, more details on $\mathcal{D}$, $\mathcal{P}$ and $\mathcal{H}$ are provided.

\paragraph{Data module $\mathcal{D}$}
The data module plays the role of Eq.~\eqref{eq:sisr} which is the closed-form solution of the data subproblem. Intuitively, it aims to find a clearer HR image which minimizes a weighted combination of the data term $\|\mathbf{y}-(\mathbf{z}\otimes \mathbf{k})\!\downarrow_{\bf{s}}\|^2$ and the quadratic regularization term $\|\mathbf{z} -\mathbf{x}_{k-1}\|^2$ with trade-off hyper-parameter $\alpha_k$.
Because the data term corresponds to the degradation model, the data module thus not only has the advantage of taking the scale factor $\bf{s}$ and blur kernel $\mathbf{k}$ as input but also imposes a degradation constraint on the solution. Actually, it is difficult to manually design such a simple but useful multiple-input module.
For brevity, Eq.~\eqref{eq:sisr} is rewritten as
\begin{equation}\label{eq:dataterm}
  \mathbf{z}_{k} = \mathcal{D}(\mathbf{x}_{k-1}, \mathbf{s}, \mathbf{k}, \mathbf{y}, \alpha_k).
\end{equation}
Note that $\mathbf{x}_0$ is initialized by interpolating $\mathbf{y}$ with scale factor $\bf{s}$ via the simplest nearest neighbor interpolation.
It should be noted that Eq.~\eqref{eq:dataterm} contains no trainable parameters, which in turn results in better generalizability due to the complete decoupling between data term and prior term.
For the implementation, we use PyTorch where the main FFT and inverse FFT operators can be implemented by \texttt{torch.rfft} and \texttt{torch.irfft}, respectively.

\paragraph{Prior module $\mathcal{P}$}
The prior module aims to obtain a cleaner HR image $\mathbf{x}_k$ by passing $\mathbf{z}_k$ through a denoiser with noise level $\beta_k$.
Inspired by~\cite{zhang2018ffdnet}, we propose a deep CNN denoiser that takes the noise level as input
\begin{equation}\label{eq:priorterm}
  \mathbf{x}_{k} = \mathcal{P}(\mathbf{z}_{k}, \beta_k).
\end{equation}
The proposed denoiser, namely ResUNet, integrates residual blocks~\cite{he2016deep} into U-Net~\cite{ronneberger2015u}.
U-Net is widely used for image-to-image mapping, while ResNet owes its popularity to fast training and its large capacity with many residual blocks.
ResUNet takes the concatenated $\mathbf{z}_{k}$ and noise level map as input and outputs the denoised image $\mathbf{x}_{k}$. By doing so, ResUNet can handle various noise levels via a single model, which significantly reduces the total number of parameters. Following the common setting of U-Net, ResUNet involves four scales, each of which has an identity skip connection between downscaling and upscaling operations.
Specifically, the number of channels in each layer from the first scale to the fourth scale are set to 64, 128, 256 and 512, respectively. For the downscaling and upscaling operations, 2$\times$2 strided convolution (SConv) and 2$\times$2 transposed convolution (TConv) are adopted, respectively.
Note that no activation function is followed by SConv and TConv layers, as well as the first and the last convolutional layers.
For the sake of inheriting the merits of ResNet, a group of 2 residual blocks are adopted in the downscaling and upscaling of each scale. As suggested in~\cite{lim2017enhanced}, each residual block is composed of two 3$\times$3 convolution layers with ReLU activation in the middle and an identity skip connection summed to its output.

\paragraph{Hyper-parameter module $\mathcal{H}$}
The hyper-parameter module acts as a `slide bar' to control the outputs of the data module and prior module.
For example, the solution $\mathbf{z}_k$ would gradually approach $\mathbf{x}_{k-1}$ as $\alpha_k$ increases.
According to the definition of $\alpha_k$ and $\beta_k$, $\alpha_k$ is determined by $\sigma$ and $\mu_k$, while $\beta_k$ depends on $\lambda$ and $\mu_k$. Although it is possible to learn a fixed $\lambda$ and $\mu_k$, we argue that a performance gain can be obtained if $\lambda$ and $\mu_k$ vary with two key elements, \ie, scale factor $\bf{s}$ and noise level $\sigma$, that influence the degree of ill-posedness.
Let $\bm{\alpha}=[\alpha_1, \alpha_2, \ldots, \alpha_K]$ and $\bm{\beta}=[\beta_1, \beta_2, \ldots, \beta_K]$,
we use a single module to predict $\bm{\alpha}$ and $\bm{\beta}$
\begin{equation}\label{eq:parameter}
  [\bm{\alpha}, \bm{\beta}] = \mathcal{H}(\sigma, \bf{s}).
\end{equation}
The hyper-parameter module consists of three fully connected layers with ReLU as the first two activation functions and Softplus~\cite{glorot2011deep} as the last.
The number of  hidden nodes in each layer is 64. Considering the fact that $\alpha_k$ and $\beta_k$ should be positive, and Eq.~\eqref{eq:sisr} should avoid division by extremely small $\alpha_k$,
the output Softplus layer is followed by an extra addition of \texttt{1e-6}.
We will show how the scale factor and noise level affect the hyper-parameters in Sec.~\ref{ssc:analysis_H}.

\subsection{End-to-end training}
\label{ssc:endtoend_training}
The end-to-end training aims to learn the trainable parameters of USRNet by minimizing a loss function over a large training data set. Thus, this section mainly describe the training data, loss function and training settings. Following~\cite{wang2018esrgan}, we use DIV2K~\cite{agustsson2017ntire} and Flickr2K~\cite{timofte2017ntire} as the HR training dataset.
The LR images are synthesized via Eq.~\eqref{eq:sisr_degradation}. Although USRNet focuses on SISR, it is also applicable to the case of deblurring with $\mathbf{s}= 1$. Hence, the scale factors are chosen from $\{1, 2, 3, 4\}$. However, due to limited space, this paper does not consider the deblurring experiments.   For the blur kernels, we use anisotropic Gaussian kernels as in~\cite{riegler2015conditioned,shocher2018zero,zhang2018learning} and motion kernels as in~\cite{boracchi2012modeling}. We fix the kernel size to $25\times25$. For the noise level, we set its range to $[0, 25]$.

With regard to the loss function, we adopt the L1 loss for PSNR performance. Following~\cite{wang2018esrgan}, once the model is obtained, we further adopt a weighted combination of L1 loss, VGG perceptual loss and relativistic adversarial loss~\cite{jolicoeur2018relativistic} with weights $1$, $1$ and $0.005$ for perceptual quality performance. We refer to such fine-tuned model as USRGAN. As usual, USRGAN only considers scale factor 4. We do not use additional losses to constrain the intermediate outputs since the above losses work well. One possible reason is that the prior module shares parameters across iterations.

To optimize the parameters of USRNet, we adopt the Adam solver~\cite{kingma2014adam} with mini-batch size 128.
The learning rate starts from $1\times10^{-4}$ and decays by a factor of 0.5 every $4\times10^4$ iterations and finally ends with $3\times10^{-6}$.
It is worth pointing out that due to the infeasibility of parallel computing for different scale factors, each min-batch only involves one random scale factor.
For USRGAN, its learning rate is fixed to $1\times10^{-5}$.
The patch size of the HR image for both USRNet and USRGAN is set to $96\times96$.
We train the models with PyTorch on 4 Nvidia Tesla V100 GPUs in Amazon AWS
cloud. It takes about two days to obtain the USRNet model.

\begin{table*}[!htbp]\scriptsize
\caption{Average PSNR(dB) results of different methods for different combinations of scale factors, blur kernels and noise levels. The best two results are highlighted in \textcolor[rgb]{1.00,0.00,0.00}{red} and \textcolor[rgb]{0.00,0.00,1.00}{blue} colors, respectively.} \vspace{-0.2cm}
\center
\begin{tabular}{p{1.1cm}<{\centering}|p{0.70cm}<{\centering}|p{0.70cm}<{\centering}|p{0.7cm}<{\centering}|p{0.7cm}<{\centering}|p{0.7cm}<{\centering}|p{0.7cm}<{\centering}|p{0.7cm}<{\centering}|p{0.7cm}<{\centering}|p{0.7cm}<{\centering}|p{0.7cm}<{\centering}|p{0.7cm}<{\centering}|p{0.7cm}<{\centering}|p{0.7cm}<{\centering}|p{0.7cm}<{\centering}}
  \hline
  \multirow{4}{*}{Method} &  & &\multicolumn{12}{c}{Blur Kernel} \\ \cline{4-15}

  & Scale & Noise & \multirow{3}{*}{\vspace{-0.1cm}\includegraphics[width=0.035\textwidth]{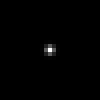}} & \multirow{3}{*}{\vspace{-0.1cm}\includegraphics[width=0.035\textwidth]{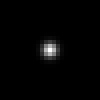}} & \multirow{3}{*}{\vspace{-0.1cm}\includegraphics[width=0.035\textwidth]{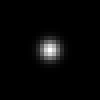}} & \multirow{3}{*}{\vspace{-0.1cm}\includegraphics[width=0.035\textwidth]{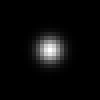}} & \multirow{3}{*}{\vspace{-0.1cm}\includegraphics[width=0.035\textwidth]{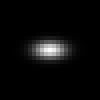}} & \multirow{3}{*}{\vspace{-0.1cm}\includegraphics[width=0.035\textwidth]{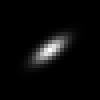}} & \multirow{3}{*}{\vspace{-0.1cm}\includegraphics[width=0.035\textwidth]{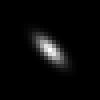}} & \multirow{3}{*}{\vspace{-0.1cm}\includegraphics[width=0.035\textwidth]{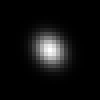}} & \multirow{3}{*}{\vspace{-0.1cm}\includegraphics[width=0.035\textwidth]{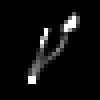}} & \multirow{3}{*}{\vspace{-0.1cm}\includegraphics[width=0.035\textwidth]{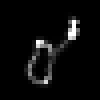}} & \multirow{3}{*}{\vspace{-0.1cm}\includegraphics[width=0.035\textwidth]{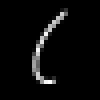}} & \multirow{3}{*}{\vspace{-0.1cm}\includegraphics[width=0.035\textwidth]{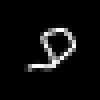}}\\
   & Factor & Level &   &  &   &   &  &  &    &  &  &  &    &  \\
    &  &  &   &  &   &   &  &  &    &  &  &  &    &  \\ \hline\hline

  & $\times$2 &0& 29.48 & 26.76 & 25.31 & 24.37 & 24.38 & 24.10  & 24.25 & 23.63 & 20.31 & 20.45 & 20.57 & 22.04\\

RCAN~\cite{zhang2018image}&$\times$3&0& 24.93&\textcolor{blue}{27.30} & 25.79 & 24.61& 24.57 & 24.38 & 24.55 & 23.74 & 20.15 & 20.25 & 20.39 & 21.68\\

& $\times$4 &0& 22.68 & \textcolor{blue}{25.31} & 25.59 & 24.63 & 24.37 & 24.23  & 24.43 & 23.74 & 20.06 & 20.05 & 20.33 & 21.47\\

\hline

  & $\times$2 &0& 29.44 & 29.48 & 28.57 & 27.42 & 27.15 & 26.81  & 27.09 & 26.25 & 14.22 & 14.22 & 16.02 & 19.39\\

ZSSR~\cite{shocher2018zero}&$\times$3&0& 25.13&25.80 & 25.94 & 25.77& 25.61 & 25.23 & 25.68 & 25.41 & 16.37 & 15.95 & 17.35 &20.45 \\

& $\times$4 &0&23.50&24.33&24.56& 24.65&24.52&24.20&24.56 &24.55& 16.94 & 16.43 &18.01&20.68\\
\hline

IKC~\cite{gu2019blind}& $\times$4 &0& 22.69 & 25.26 & \textcolor{blue}{25.63} & 25.21 & 24.71 & 24.20  & 24.39 & 24.77 & 20.05 & 20.03 & 20.35 & 21.58\\
\hline

  & $\times$2 &0& \textcolor{blue}{29.60}& \textcolor{blue}{30.16} & \textcolor{blue}{29.50} &\textcolor{blue}{28.37} & \textcolor{blue}{28.07} & \textcolor{blue}{27.95} &\textcolor{blue}{28.21} &\textcolor{blue}{27.19}& \textcolor{blue}{28.58}&\textcolor{blue}{26.79} &\textcolor{blue}{29.02} & \textcolor{blue}{28.96}\\

  & $\times$3 &0& \textcolor{blue}{25.97}& 26.89 & \textcolor{blue}{27.07}& \textcolor{blue}{27.01}& \textcolor{blue}{26.83} &\textcolor{blue}{26.76} &\textcolor{blue}{26.88}& \textcolor{blue}{26.67} &\textcolor{blue}{26.22} &\textcolor{blue}{25.59} &\textcolor{blue}{26.14}  &\textcolor{blue}{26.05}\\

IRCNN~\cite{zhang2017learning}&$\times$3&2.55& \textcolor{blue}{25.70} &\textcolor{blue}{26.13}& \textcolor{blue}{25.72}& \textcolor{blue}{25.33} &\textcolor{blue}{25.28} &\textcolor{blue}{25.18}& \textcolor{blue}{25.34}& \textcolor{blue}{24.97}& \textcolor{blue}{25.00}& \textcolor{blue}{24.64}& \textcolor{blue}{24.90}& \textcolor{blue}{24.73}\\
  & $\times$3 & 7.65 & \textcolor{blue}{24.58} &\textcolor{blue}{24.68} &\textcolor{blue}{24.59}& \textcolor{blue}{24.39}& \textcolor{blue}{24.24}& \textcolor{blue}{24.20}& \textcolor{blue}{24.27} &\textcolor{blue}{24.02} &\textcolor{blue}{23.94} &\textcolor{blue}{23.77}& \textcolor{blue}{23.75} & \textcolor{blue}{23.69}\\

& $\times$4 &0& \textcolor{blue}{23.99} &25.01 &25.32 &\textcolor{blue}{25.45} &\textcolor{blue}{25.36} &\textcolor{blue}{25.26} &\textcolor{blue}{25.34}& \textcolor{blue}{25.47} &\textcolor{blue}{24.69} &\textcolor{blue}{24.39}& \textcolor{blue}{24.44} &\textcolor{blue}{24.57}\\
\hline

  & $\times$2 &0& \textcolor{red}{30.55} & \textcolor{red}{30.96} & \textcolor{red}{30.56} & \textcolor{red}{29.49} & \textcolor{red}{29.13} & \textcolor{red}{29.12} & \textcolor{red}{29.28} & \textcolor{red}{28.28} & \textcolor{red}{30.90} & \textcolor{red}{30.65} & \textcolor{red}{30.60} & \textcolor{red}{30.75}\\

  & $\times$3 &0& \textcolor{red}{27.16} & \textcolor{red}{27.76} & \textcolor{red}{27.90} & \textcolor{red}{27.88} & \textcolor{red}{27.71} & \textcolor{red}{27.68} & \textcolor{red}{27.74} & \textcolor{red}{27.57} & \textcolor{red}{27.69} & \textcolor{red}{27.50} & \textcolor{red}{27.50} & \textcolor{red}{27.41}\\

USRNet&$\times$3&2.55& \textcolor{red}{26.99}&\textcolor{red}{27.40} & \textcolor{red}{27.23} & \textcolor{red}{26.78}& \textcolor{red}{26.55} & \textcolor{red}{26.60} & \textcolor{red}{26.72} & \textcolor{red}{26.14} & \textcolor{red}{26.90} &\textcolor{red}{26.80} & \textcolor{red}{26.69} & \textcolor{red}{26.49}\\

  & $\times$3 &7.65& \textcolor{red}{26.45} & \textcolor{red}{26.52} & \textcolor{red}{26.10} & \textcolor{red}{25.57} & \textcolor{red}{25.46} & \textcolor{red}{25.40} & \textcolor{red}{25.49} & \textcolor{red}{25.00} & \textcolor{red}{25.39} & \textcolor{red}{25.47} & \textcolor{red}{25.20} & \textcolor{red}{25.01}\\

& $\times$4 &0& \textcolor{red}{25.30} & \textcolor{red}{25.96} & \textcolor{red}{26.18} & \textcolor{red}{26.29} & \textcolor{red}{26.20} & \textcolor{red}{26.15} & \textcolor{red}{26.17} & \textcolor{red}{26.30} & \textcolor{red}{25.91} & \textcolor{red}{25.57} & \textcolor{red}{25.76} & \textcolor{red}{25.70}\\
\hline

\end{tabular}
\label{table:psnr}\vspace{-0.05cm}
\end{table*}

\begin{figure*}[!htbp]\footnotesize 
\hspace{-0.26cm}
\begin{tabular}{c@{\extracolsep{0em}}|@{\extracolsep{0.25em}}c@{\extracolsep{0.05em}}c@{\extracolsep{0.05em}}c@{\extracolsep{0.05em}}c@{\extracolsep{0.00em}}|@{\extracolsep{0.25em}}c@{\extracolsep{0.05em}}c}
        \includegraphics[width=0.137\textwidth]{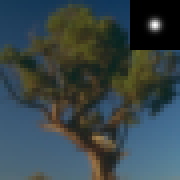}~
		&\includegraphics[width=0.137\textwidth]{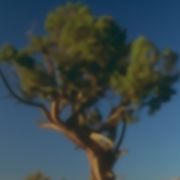}~
		&\includegraphics[width=0.137\textwidth]{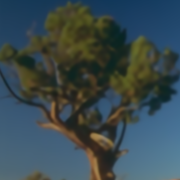}~
        &\includegraphics[width=0.137\textwidth]{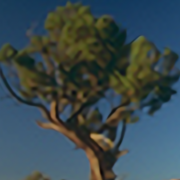}~
        &\includegraphics[width=0.137\textwidth]{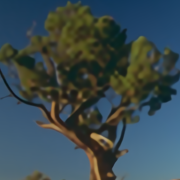}~
		&\includegraphics[width=0.137\textwidth]{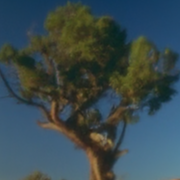}~
		&\includegraphics[width=0.137\textwidth]{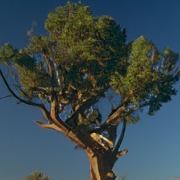}\\
        \includegraphics[width=0.137\textwidth]{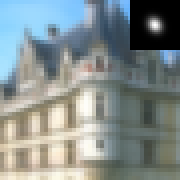}~
		&\includegraphics[width=0.137\textwidth]{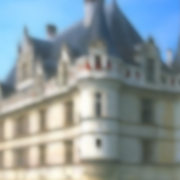}~
		&\includegraphics[width=0.137\textwidth]{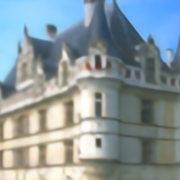}~
        &\includegraphics[width=0.137\textwidth]{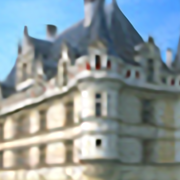}~
        &\includegraphics[width=0.137\textwidth]{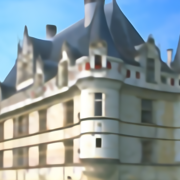}~
		&\includegraphics[width=0.137\textwidth]{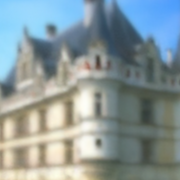}~
		&\includegraphics[width=0.137\textwidth]{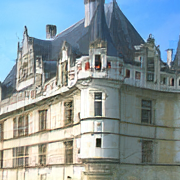}\\

        \includegraphics[width=0.137\textwidth]{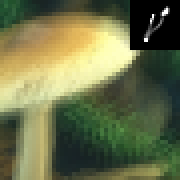}~
		&\includegraphics[width=0.137\textwidth]{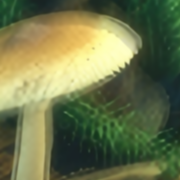}~
		&\includegraphics[width=0.137\textwidth]{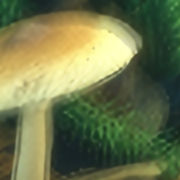}~
        &\includegraphics[width=0.137\textwidth]{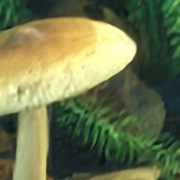}~
        &\includegraphics[width=0.137\textwidth]{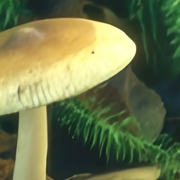}~
		&\includegraphics[width=0.137\textwidth]{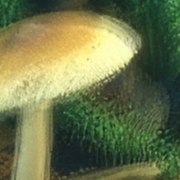}~
		&\includegraphics[width=0.137\textwidth]{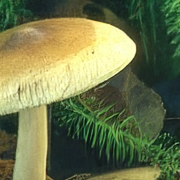}\\
  PSNR(dB) & 24.92/22.53/21.88 & 25.24/23.75/21.92 & 25.42/23.55/26.45 &25.82/24.80/27.39 &24.80/22.44/21.51 &24.26/23.25/25.00  \\

Zoomed LR ($\times$4) &RCAN~\cite{zhang2018image} & IKC~\cite{gu2019blind}  & IRCNN~\cite{zhang2017learning} & \textbf{USRNet} (ours) & RankSRGAN~\cite{zhang2019ranksrgan} & \textbf{USRGAN} (ours) \\
	\end{tabular}
    \vspace{0.1cm}
	\caption{Visual results of different methods on super-resolving noise-free LR image with scale factor 4. The blur kernel is shown on the upper-right corner of the LR image. Note that RankSRGAN and our USRGAN aim for perceptual quality rather than PSNR value.}
	\label{fig:visualresults}\vspace{-0.1cm}
\end{figure*}

\section{Experiments}\label{sec:experiments}
We choose the widely-used color BSD68 dataset~\cite{MartinFTM01,roth2009fields} to quantitatively evaluate different methods. The dataset consists of 68 images with tiny structures and fine textures and thus is challenging to improve the quantitative metrics, such as PSNR.
For the sake of synthesizing the corresponding testing LR images via Eq.~\eqref{eq:sisr_degradation}, blur kernels and noise levels should be provided. Generally, it would be helpful to employ a large variety of blur kernels and noise levels for a thorough evaluation, however, it would also give rise to burdensome evaluation process. For this reason, as shown in Table~\ref{table:psnr}, we only consider 12 representative and diverse blur kernels, including 4 isotropic Gaussian kernels with different widths (\ie, 0.7, 1.2, 1.6 and 2.0), 4 anisotropic Gaussian kernels from~\cite{zhang2018learning}, and 4 motion blur kernels from ~\cite{levin2009understanding,boracchi2012modeling}.
While it has been pointed out that anisotropic Gaussian kernels are enough for SISR task~\cite{riegler2015conditioned,shocher2018zero}, the SISR method that can handle more complex blur kernels would be a preferred choice in real applications.
Therefore, it is necessary to further analyze the kernel robustness of different methods, we will thus separately report the PSNR results for each blur kernel rather than for each type of blur kernels.
Although it has been pointed out that the proper blur kernel should vary with scale factor~\cite{zhang2015revisiting}, we argue that the 12 blur kernels are diverse enough to cover a large kernel space.
For the noise levels, we choose 2.55 (1\%) and 7.65 (3\%).

\begin{figure*}[!htbp]\footnotesize
\hspace{-0.22cm}
\begin{tabular}{c@{\extracolsep{0.00em}}|@{\extracolsep{0.25em}}c@{\extracolsep{0.00em}}c@{\extracolsep{0.00em}}c@{\extracolsep{0.00em}}|@{\extracolsep{0.25em}}c@{\extracolsep{0.00em}}c@{\extracolsep{0.00em}}c}
        \includegraphics[width=0.137\textwidth]{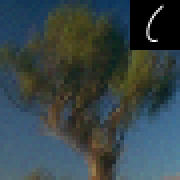}~
		&\includegraphics[width=0.137\textwidth]{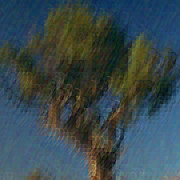}~
		&\includegraphics[width=0.137\textwidth]{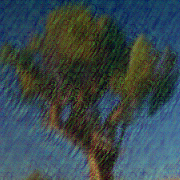}~
        &\includegraphics[width=0.137\textwidth]{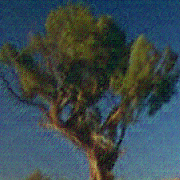}~
        &\includegraphics[width=0.137\textwidth]{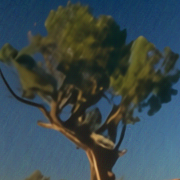}~
		&\includegraphics[width=0.137\textwidth]{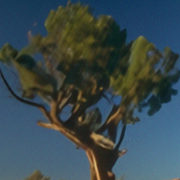}~
		&\includegraphics[width=0.137\textwidth]{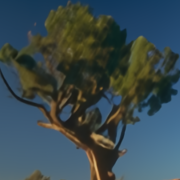}\\

        \includegraphics[width=0.137\textwidth]{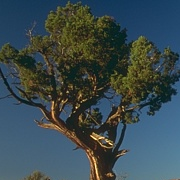}~
		&\includegraphics[width=0.137\textwidth]{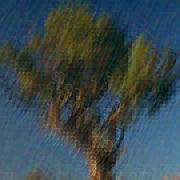}~
		&\includegraphics[width=0.137\textwidth]{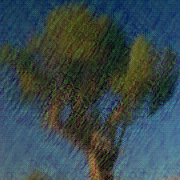}~
        &\includegraphics[width=0.137\textwidth]{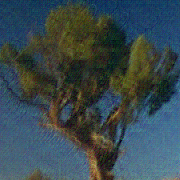}~
        &\includegraphics[width=0.137\textwidth]{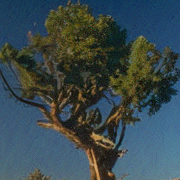}~
		&\includegraphics[width=0.137\textwidth]{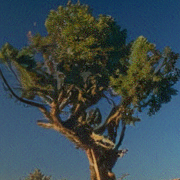}~
		&\includegraphics[width=0.137\textwidth]{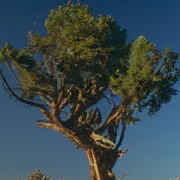}\\

$\mathbf{x}_0$ (Top); $\mathbf{x}$ (Bottom) & \multicolumn{3}{c}{$\mathbf{z}_1\qquad\;\; \longrightarrow\qquad\;\;~\mathbf{x}_1\qquad\;\;\longrightarrow\qquad\;\;\mathbf{z}_2$} &  \multicolumn{3}{c}{$\mathbf{x}_7\qquad\;\;\longrightarrow\qquad\;\;~\mathbf{z}_8\qquad\;\; \longrightarrow\qquad\;\;\mathbf{x}_8$}~~\\

	\end{tabular}
    \vspace{-0.04cm}
	\caption{HR estimations in different iterations of USRNet (top row) and USRGAN (bottom row). The initial HR estimation $\mathbf{x}_0$ is the nearest neighbor interpolated version of LR image. The scale factor is 4, the noise level of LR image is 2.55 (1\%), the blur kernel is shown on the upper-right corner of $\mathbf{x}_0$.}
	\label{fig:output_DP}\vspace{-0.1cm}
\end{figure*}

\subsection{PSNR results}
The average PSNR results of different methods for different degradation settings are reported in Table~\ref{table:psnr}.
The compared methods include RCAN~\cite{zhang2018image}, ZSSR~\cite{shocher2018zero}, IKC~\cite{gu2019blind} and IRCNN~\cite{zhang2017learning}. Specifically, RCAN is state-of-the-art PSNR oriented method for bicubic degradation; ZSSR is a non-blind zero-shot learning method with the ability to handle Eq.~\eqref{eq:sisr_degradation} for anisotropic Gaussian kernels; IKC is a blind iterative kernel correction method for isotropic Gaussian kernels; IRCNN a non-blind deep denoiser based plug-and-play method.
For a fair comparison, we modified IRCNN to handle Eq.~\eqref{eq:sisr_degradation} by replacing its data solution with Eq.~\eqref{eq:sisr}.
Note that following~\cite{liu2013bayesian}, we fix the pixel shift issue before calculating PSNR if necessary.

According to Table~\ref{table:psnr}, we can have the following observations. First, our USRNet with a single model significantly outperforms the other competitive methods on different scale factors, blur kernels and noise levels. In particular, with much fewer iterations, USRNet has at least an average PSNR gain of 1dB over IRCNN with 30 iterations due to the end-to-end training.
Second, RCAN can achieve good performance on the degradation setting similar to bicubic degradation but would deteriorate seriously when the degradation deviates from bicubic degradation. Such a phenomenon has been well studied in~\cite{efrat2013accurate}.
Third, ZSSR performs well on both isotropic and anisotropic Gaussian blur kernels for small scale factors but loses effectiveness on motion blur kernel and large scale factors. Actually, ZSSR has difficulty in capturing natural image characteristic on severely degraded image due to the single image learning strategy.
Fourth, IKC does not generalize well to anisotropic Gaussian kernels and motion kernels.

Although USRNet is not designed for bicubic degradation, it is interesting to test its results by taking the approximated bicubic kernels in Fig.~\ref{fig:bicubic_kernel} as input.
From Table~\ref{table_bicubic}, one can see that USRNet still performs favorably without training on the bicubic kernels.

\begin{table}[!htbp]\footnotesize
\caption{The average PSNR(dB) results of USRNet for bicubic degradation on commonly-used testing datasets.}
\center
\begin{tabular}{|p{1.7cm}<{\centering}|p{1.1cm}<{\centering}|p{1.1cm}<{\centering}|p{1.1cm}<{\centering}|p{1.1cm}<{\centering}|}
  \hline
  Scale Factor& Set5& Set14   & BSD100 & Urban100 \\ \hline
  $\times$2 &  37.72& 33.49  &  32.10 & 31.79  \\\hline
   $\times$3 &  34.45  & 30.51 & 29.18   & 28.38\\\hline
   $\times$4 &  32.45  & 28.83 &  27.69 & 26.44 \\
  \hline
\end{tabular}
\label{table_bicubic}
\end{table}

\subsection{Visual results}
The visual results of different methods on super-resolving noise-free LR image with scale factor 4 are shown in Fig.~\ref{fig:visualresults}. Apart from RCAN, IKC and IRCNN, we also include RankSRGAN~\cite{zhang2019ranksrgan} for comparison with our USRGAN. Note that the visual results of ZSSR are omitted due to the inferior performance on scale factor 4. It can be observed from Fig.~\ref{fig:visualresults} that USRNet and IRCNN produce much better visual results than RCAN and IKC on the LR image with motion blur kernel.
While USRNet can recover shaper edges than IRCNN, both of them fail to produce realistic textures.
As expected, USRGAN can yield much better visually pleasant results than USRNet. On the other hand, RankSRGAN
does not perform well if the degradation largely deviates from the bicubic degradation.
In contrast, USRGAN is flexible to handle various LR images.

\subsection{Analysis on $\mathcal{D}$ and $\mathcal{P}$}
\label{ssc:analysis_DP}
Because the proposed USRNet is an iterative method, it is interesting to investigate the HR estimations of data module $\mathcal{D}$ and prior module $\mathcal{P}$ in different iterations.
Fig.~\ref{fig:output_DP} shows the results of USRNet and USRGAN in different iterations for an LR image with scale factor 4.
As one can see, $\mathcal{D}$ and $\mathcal{P}$ can facilitate each other for iterative and alternating blur removal and detail recovery.
Interestingly, $\mathcal{P}$ can also act as a detail enhancer for high-frequency recovery due to the task-specific training. In addition, it does not reduce blur kernel induced degradation which verifies the decoupling between $\mathcal{D}$ and $\mathcal{P}$.
As a result, the end-to-end trained USRNet has a task-specific advantage over Gaussian denoiser based plug-and-play SISR.
To quantitatively analyze the role of $\mathcal{D}$, we have trained an USRNet model with 5 iterations, it turns out that the average PSNR value will decreases about 0.1dB on Gaussian blur kernels and 0.3dB on motion blur kernels. This further indicates that $\mathcal{D}$ aims to eliminate blur kernel induced degradation.
In addition, one can see that USRGAN has similar results with USRNet in the first few iterations, but will instead recover tiny structures and fine textures in last few iterations.

\subsection{Analysis on $\mathcal{H}$}
\label{ssc:analysis_H}

\begin{figure}[!htbp]\footnotesize
\hspace{-0.23cm}
	\begin{tabular}{c@{\extracolsep{0.01em}}c@{\extracolsep{0.01em}}}
		\includegraphics[width=0.23\textwidth]{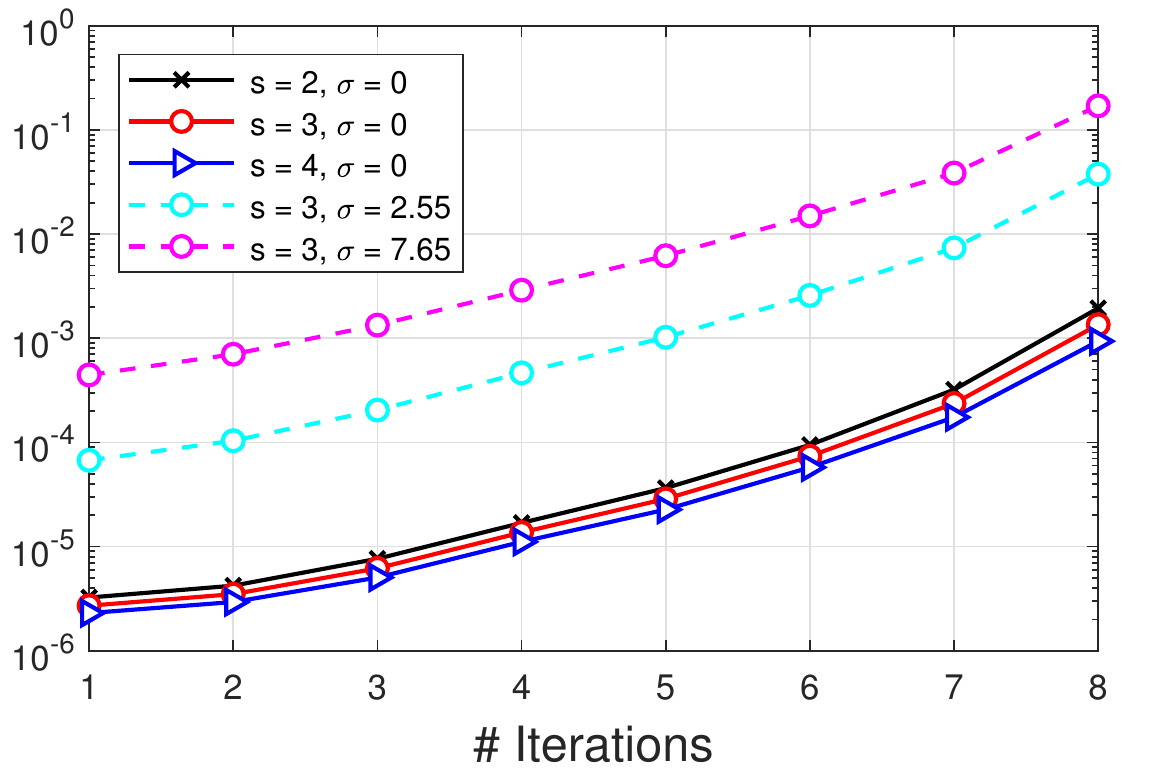}~
		&\includegraphics[width=0.23\textwidth]{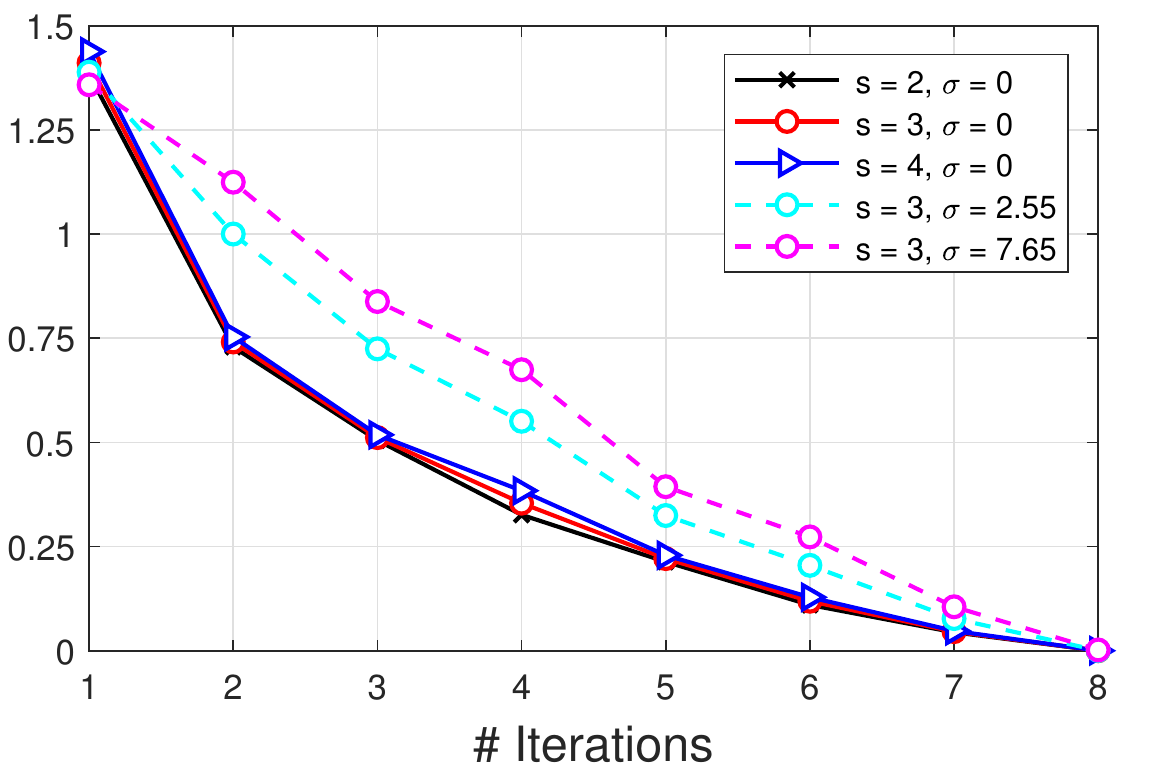}\\
		(a) $\bm{\alpha}$ & (b) $\bm{\beta}$ \\
	\end{tabular}\vspace{0.001cm}
	\caption{Outputs of the hyper-parameter module $\mathcal{H}$, \ie, (a) $\bm{\alpha}$ and (b) $\bm{\beta}$,  with respect to different combinations of $\bf{s}$ and $\sigma$.}
	\label{fig:analysis_H}\vspace{-0.2cm}
\end{figure}

Fig.~\ref{fig:analysis_H} shows outputs of the hyper-parameter module for different combinations of scale factor $\bf{s}$ and noise level $\sigma$. It can be observed from Fig.~\ref{fig:analysis_H}(a) that
$\bm{\alpha}$ is positively correlated with $\sigma$ and varies with $\mathbf{s}$. This actually accords with the definition of  $\alpha_i$ in Sec.~\ref{section:deepunfolding} and our analysis in Sec.~\ref{section:deepnetwork}. From Fig.~\ref{fig:analysis_H}(b), one can see that
$\bm{\beta}$ has a decreasing tendency with the number of iterations and increases with scale factor and noise level. This implies that the noise level of HR estimation is gradually reduced across iterations and
complex degradation requires a large $\beta_i$ to tackle with the illposeness. It should be pointed out that the learned hyper-parameter setting is in accordance with that of IRCNN~\cite{zhang2017learning}.
In summary, the learned $\mathcal{H}$ is meaningful as it plays the proper role.

\subsection{Generalizability}
\vspace{-0.2cm}
\begin{figure}[!htbp]\footnotesize
\hspace{-0.18cm}
	\begin{tabular}{c@{\extracolsep{0.2em}}c@{\extracolsep{0.01em}}}
		\includegraphics[width=0.225\textwidth]{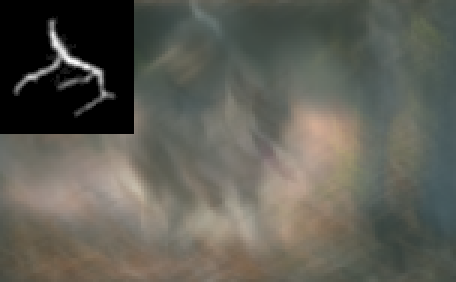}~
		&\includegraphics[width=0.225\textwidth]{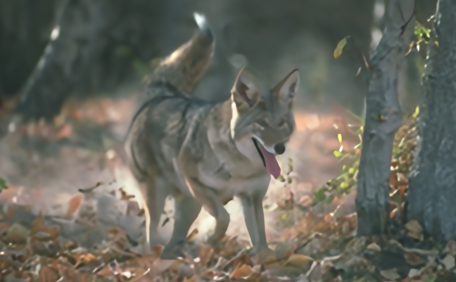}\\
(a) Zoomed LR ($\times$3) & (b) USRNet\\
        \includegraphics[width=0.225\textwidth]{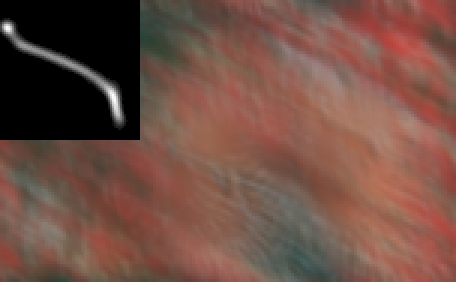}~
		&\includegraphics[width=0.225\textwidth]{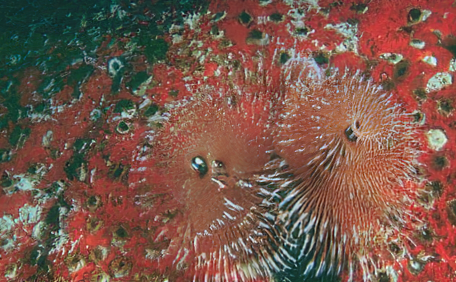}\\
(c) Zoomed LR ($\times$3) & (d) USRGAN
	\end{tabular}
\vspace{0.001cm}
	\caption{An illustration to show the generalizability of USRNet and USRGAN. The sizes of the kernels in (a) and (c) are 67$\times$67 and 70$\times$70, respectively. The two kernels are chosen from~\cite{pan2016blind}.}
	\label{fig:generalizability}\vspace{-0.2cm}
\end{figure}

As mentioned earlier, the proposed method enjoys good generalizability due to the decoupling of data term and prior term. To demonstrate such an advantage, Fig.~\ref{fig:generalizability} shows the visual results of USRNet and USRGAN on LR image with a kernel of much larger size than training size of 25$\times$25. It can be seen that both USRNet and USRGAN can produce visually pleasant results, which can be attributed to the trainable parameter-free data module.
It is worth pointing out that USRGAN is trained on scale factor 4, while Fig.~\ref{fig:generalizability}(b) shows its visual result on scale factor 3. This further indicates that the prior module of USRGAN can generalize to other scale factors. In summary, the proposed deep unfolding architecture has superiority in generalizability.

\subsection{Real image super-resolution}
\begin{figure}[!htbp]\footnotesize
\hspace{-0.18cm}
	\begin{tabular}{c@{\extracolsep{0.2em}}c@{\extracolsep{0.01em}}}
		\includegraphics[width=0.225\textwidth]{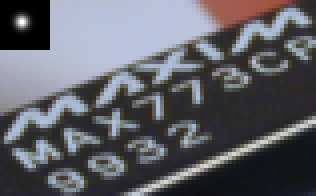}~
		&\includegraphics[width=0.225\textwidth]{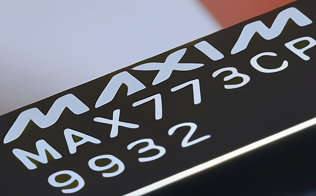}\\
		(a) Zoomed LR ($\times$4) & (b) USRNet \\
	\end{tabular}
\vspace{0.01cm}
	\caption{Visual result of USRNet ($\times4$) on a real LR image.}
	\label{fig:real}\vspace{-0.27cm}
\end{figure}

Because Eq.~\eqref{eq:sisr} is based on the assumption of circular boundary condition, a proper boundary handling for the real LR image is generally required. We use the following three steps to do such pre-processing. First, the LR image is interpolated to the desired size. Second, the boundary handling method proposed in~\cite{liu2008reducing} is adopted on the interpolated image with the blur kernel. Last, the downsampled boundaries are padded to the original LR image.
Fig.~\ref{fig:real} shows the visual result of USRNet on real LR image with scale factor 4.
The blur kernel is manually selected as isotropic Gaussian kernel with width 2.2 based on user preference.
One can see from Fig.~\ref{fig:real} that the proposed USRNet can reconstruct the HR image with improved visual quality.

\section{Conclusion}
\label{sec:conclusion}
In this paper, we focus on the classical SISR degradation model and propose a deep unfolding super-resolution network. Inspired by the unfolding optimization of traditional model-based method, we design an end-to-end trainable deep network which integrates the flexibility of model-based methods and the advantages of learning-based methods.
The main novelty of the proposed network is that it can handle the classical degradation model via a single model.
Specifically, the proposed network consists of three interpretable modules, including the data module that makes HR estimation clearer, the prior module that makes HR estimation cleaner, and the hyper-parameter module that controls the outputs of the other two modules.
As a result, the proposed method can impose both degradation constrain and prior constrain on the solution.
Extensive experimental results demonstrated the flexibility, effectiveness and generalizability of the proposed method for super-resolving various degraded LR images.
We believe that our work can benefit to image restoration research community.

\vspace{0.1cm}
\noindent\textbf{Acknowledgments:}
This work was partly supported by the ETH Z\"urich Fund (OK), a Huawei Technologies Oy (Finland) project, an Amazon AWS grant, and Nvidia.

\clearpage

{\small
\bibliographystyle{ieee_fullname}
\bibliography{egbib}

\begin{thebibliography}{10}\itemsep=-1pt

\bibitem{adler2018learned}
Jonas Adler and Ozan {\"O}ktem.
\newblock Learned primal-dual reconstruction.
\newblock {\em IEEE TMI}, 37(6):1322--1332, 2018.

\bibitem{afonso2010fast}
Manya~V Afonso, Jos{\'e}~M Bioucas-Dias, and M{\'a}rio~AT Figueiredo.
\newblock Fast image recovery using variable splitting and constrained
  optimization.
\newblock {\em IEEE TIP}, 19(9):2345--2356, 2010.

\bibitem{agustsson2017ntire}
Eirikur Agustsson and Radu Timofte.
\newblock Ntire 2017 challenge on single image super-resolution: Dataset and
  study.
\newblock In {\em CVPRW}, volume~3, pages 126--135, July 2017.

\bibitem{barbu2009training}
Adrian Barbu.
\newblock Training an active random field for real-time image denoising.
\newblock {\em IEEE TIP}, 18(11):2451--2462, 2009.

\bibitem{boracchi2012modeling}
Giacomo Boracchi and Alessandro Foi.
\newblock Modeling the performance of image restoration from motion blur.
\newblock {\em IEEE TIP}, 21(8):3502--3517, 2012.

\bibitem{boyd2011distributed}
Stephen Boyd, Neal Parikh, Eric Chu, Borja Peleato, and Jonathan Eckstein.
\newblock Distributed optimization and statistical learning via the alternating
  direction method of multipliers.
\newblock {\em Foundations and Trends in Machine Learning}, 3(1):1--122, 2011.

\bibitem{brifman2019unified}
Alon Brifman, Yaniv Romano, and Michael Elad.
\newblock Unified single-image and video super-resolution via denoising
  algorithms.
\newblock {\em IEEE TIP}, 28(12):6063--6076, 2019.

\bibitem{caselles1998axiomatic}
Vicent Caselles, J-M Morel, and Catalina Sbert.
\newblock An axiomatic approach to image interpolation.
\newblock {\em IEEE TIP}, 7(3):376--386, 1998.

\bibitem{chambolle2011first}
Antonin Chambolle and Thomas Pock.
\newblock A first-order primal-dual algorithm for convex problems with
  applications to imaging.
\newblock {\em Journal of Mathematical Imaging and Vision}, 40(1):120--145,
  2011.

\bibitem{chan2016plug}
Stanley~H Chan, Xiran Wang, and Omar~A Elgendy.
\newblock {Plug-and-Play} {ADMM} for image restoration: Fixed-point convergence
  and applications.
\newblock {\em IEEE Transactions on Computational Imaging}, 3(1):84--98, 2017.

\bibitem{chen2015trainable}
Yunjin Chen and Thomas Pock.
\newblock Trainable nonlinear reaction diffusion: {A} flexible framework for
  fast and effective image restoration.
\newblock {\em IEEE TPAMI}, 2016.

\bibitem{chen2018fsrnet}
Yu Chen, Ying Tai, Xiaoming Liu, Chunhua Shen, and Jian Yang.
\newblock Fsrnet: End-to-end learning face super-resolution with facial priors.
\newblock In {\em CVPR}, pages 2492--2501, 2018.

\bibitem{dai2016image}
Dengxin Dai, Yujian Wang, Yuhua Chen, and Luc Van~Gool.
\newblock Is image super-resolution helpful for other vision tasks?
\newblock In {\em WACV}, pages 1--9, 2016.

\bibitem{Dong2016}
Chao Dong, C.~C. Loy, Kaiming He, and Xiaoou Tang.
\newblock Image super{-}resolution using deep convolutional networks.
\newblock {\em IEEE TPAMI}, 38(2):295--307, 2016.

\bibitem{dong2013nonlocally}
Weisheng Dong, Lei Zhang, Guangming Shi, and Xin Li.
\newblock Nonlocally centralized sparse representation for image restoration.
\newblock {\em IEEE TIP}, 22(4):1620--1630, 2013.

\bibitem{efrat2013accurate}
Netalee Efrat, Daniel Glasner, Alexander Apartsin, Boaz Nadler, and Anat Levin.
\newblock Accurate blur models vs. image priors in single image
  super-resolution.
\newblock In {\em ICCV}, pages 2832--2839, 2013.

\bibitem{elad1997restoration}
Michael Elad and Arie Feuer.
\newblock Restoration of a single superresolution image from several blurred,
  noisy, and undersampled measured images.
\newblock {\em IEEE TIP}, 6(12):1646--1658, 1997.

\bibitem{farsiu2004advances}
Sina Farsiu, Dirk Robinson, Michael Elad, and Peyman Milanfar.
\newblock Advances and challenges in super-resolution.
\newblock {\em International Journal of Imaging Systems and Technology},
  14(2):47--57, 2004.

\bibitem{glorot2011deep}
Xavier Glorot, Antoine Bordes, and Yoshua Bengio.
\newblock Deep sparse rectifier neural networks.
\newblock In {\em ICAIS}, pages 315--323, 2011.

\bibitem{gu2019blind}
Jinjin Gu, Hannan Lu, Wangmeng Zuo, and Chao Dong.
\newblock Blind super-resolution with iterative kernel correction.
\newblock In {\em CVPR}, pages 1604--1613, 2019.

\bibitem{he2016deep}
Kaiming He, Xiangyu Zhang, Shaoqing Ren, and Jian Sun.
\newblock Deep residual learning for image recognition.
\newblock In {\em CVPR}, pages 770--778, 2016.

\bibitem{heide2016proximal}
Felix Heide, Steven Diamond, Matthias Niener, Jonathan Ragan-Kelley, Wolfgang
  Heidrich, and Gordon Wetzstein.
\newblock Proximal: Efficient image optimization using proximal algorithms.
\newblock {\em ACM TOG}, 35(4):84, 2016.

\bibitem{hu2019meta}
Xuecai Hu, Haoyuan Mu, Xiangyu Zhang, Zilei Wang, Tieniu Tan, and Jian Sun.
\newblock {Meta-SR}: A magnification-arbitrary network for super-resolution.
\newblock In {\em CVPR}, pages 1575--1584, 2019.

\bibitem{jolicoeur2018relativistic}
Alexia Jolicoeur-Martineau.
\newblock The relativistic discriminator: a key element missing from standard
  {GAN}.
\newblock {\em arXiv:1807.00734}, 2018.

\bibitem{keys1981cubic}
Robert Keys.
\newblock Cubic convolution interpolation for digital image processing.
\newblock {\em IEEE Transactions on Acoustics, Speech, and Signal Processing},
  29(6):1153--1160, 1981.

\bibitem{kim2015accurate}
Jiwon Kim, Jung~Kwon Lee, and Kyoung~Mu Lee.
\newblock Accurate image super-resolution using very deep convolutional
  networks.
\newblock In {\em CVPR}, pages 1646--1654, 2016.

\bibitem{kingma2014adam}
Diederik Kingma and Jimmy Ba.
\newblock Adam: A method for stochastic optimization.
\newblock In {\em ICLR}, 2015.

\bibitem{kokkinos2018deep}
Filippos Kokkinos and Stamatios Lefkimmiatis.
\newblock Deep image demosaicking using a cascade of convolutional residual
  denoising networks.
\newblock In {\em ECCV}, pages 303--319, 2018.

\bibitem{kruse2017learning}
Jakob Kruse, Carsten Rother, and Uwe Schmidt.
\newblock Learning to push the limits of efficient {FFT}-based image
  deconvolution.
\newblock In {\em ICCV}, pages 4586--4594, 2017.

\bibitem{lai2017deep}
Wei-Sheng Lai, Jia-Bin Huang, Narendra Ahuja, and Ming-Hsuan Yang.
\newblock Deep laplacian pyramid networks for fast and accurate
  super-resolution.
\newblock In {\em CVPR}, pages 624--632, July 2017.

\bibitem{ledig2016photo}
Christian Ledig, Lucas Theis, Ferenc Husz{\'a}r, Jose Caballero, Andrew
  Cunningham, Alejandro Acosta, Andrew Aitken, Alykhan Tejani, Johannes Totz,
  Zehan Wang, et~al.
\newblock Photo-realistic single image super-resolution using a generative
  adversarial network.
\newblock In {\em CVPR}, pages 4681--4690, July 2017.

\bibitem{lefkimmiatis2016non}
Stamatios Lefkimmiatis.
\newblock Non-local color image denoising with convolutional neural networks.
\newblock In {\em CVPR}, pages 3587--3596, 2017.

\bibitem{levin2009understanding}
Anat Levin, Yair Weiss, Fredo Durand, and William~T Freeman.
\newblock Understanding and evaluating blind deconvolution algorithms.
\newblock In {\em CVPR}, pages 1964--1971, 2009.

\bibitem{li2017iterative}
Tao Li, Xiaohai He, Linbo Qing, Qizhi Teng, and Honggang Chen.
\newblock An iterative framework of cascaded deblocking and superresolution for
  compressed images.
\newblock {\em IEEE Transactions on Multimedia}, 20(6):1305--1320, 2017.

\bibitem{li2020group}
Yawei Li, Shuhang Gu, Christoph Mayer, Luc Van~Gool, and Radu Timofte.
\newblock Group sparsity: The hinge between filter pruning and decomposition
  for network compression.
\newblock In {\em CVPR}, 2020.

\bibitem{lim2017enhanced}
Bee Lim, Sanghyun Son, Heewon Kim, Seungjun Nah, and Kyoung~Mu Lee.
\newblock Enhanced deep residual networks for single image super-resolution.
\newblock In {\em CVPRW}, pages 136--144, July 2017.

\bibitem{liu2013bayesian}
Ce Liu and Deqing Sun.
\newblock On bayesian adaptive video super resolution.
\newblock {\em IEEE TPAMI}, 36(2):346--360, 2013.

\bibitem{liu2008reducing}
Renting Liu and Jiaya Jia.
\newblock Reducing boundary artifacts in image deconvolution.
\newblock In {\em ICIP}, pages 505--508, 2008.

\bibitem{lugmayr2019unsupervised}
Andreas {Lugmayr}, Martin {Danelljan}, and Radu {Timofte}.
\newblock Unsupervised learning for real-world super-resolution.
\newblock In {\em ICCVW}, pages 3408--3416, 2019.

\bibitem{MartinFTM01}
D. Martin, C. Fowlkes, D. Tal, and J. Malik.
\newblock A database of human segmented natural images and its application to
  evaluating segmentation algorithms and measuring ecological statistics.
\newblock In {\em ICCV}, pages 416--423, 2001.

\bibitem{pan2016blind}
Jinshan Pan, Deqing Sun, Hanspeter Pfister, and Ming-Hsuan Yang.
\newblock Blind image deblurring using dark channel prior.
\newblock In {\em CVPR}, pages 1628--1636, 2016.

\bibitem{peleg2014statistical}
Tomer Peleg and Michael Elad.
\newblock A statistical prediction model based on sparse representations for
  single image super-resolution.
\newblock {\em IEEE TIP}, 23(6):2569--2582, 2014.

\bibitem{ren2019neural}
Dongwei Ren, Kai Zhang, Qilong Wang, Qinghua Hu, and Wangmeng Zuo.
\newblock Neural blind deconvolution using deep priors.
\newblock In {\em CVPR}, pages 1628--1636, 2020.

\bibitem{riegler2015conditioned}
Gernot Riegler, Samuel Schulter, Matthias Ruther, and Horst Bischof.
\newblock Conditioned regression models for non-blind single image
  super-resolution.
\newblock In {\em ICCV}, pages 522--530, 2015.

\bibitem{ronneberger2015u}
Olaf Ronneberger, Philipp Fischer, and Thomas Brox.
\newblock U-net: Convolutional networks for biomedical image segmentation.
\newblock In {\em International Conference on Medical image computing and
  computer-assisted intervention}, pages 234--241. Springer, 2015.

\bibitem{roth2009fields}
Stefan Roth and Michael~J Black.
\newblock Fields of experts.
\newblock {\em IJCV}, 82(2):205--229, 2009.

\bibitem{sajjadi2017enhancenet}
Mehdi~SM Sajjadi, Bernhard Sch{\"o}lkopf, and Michael Hirsch.
\newblock Enhancenet: Single image super-resolution through automated texture
  synthesis.
\newblock In {\em ICCV}, pages 4501--4510, 2017.

\bibitem{samuel2009learning}
Kegan~GG Samuel and Marshall~F Tappen.
\newblock Learning optimized {MAP} estimates in continuously-valued {MRF}
  models.
\newblock In {\em CVPR}, pages 477--484, 2009.

\bibitem{schmidt2014shrinkage}
Uwe Schmidt and Stefan Roth.
\newblock Shrinkage fields for effective image restoration.
\newblock In {\em CVPR}, pages 2774--2781, 2014.

\bibitem{shen2018deep}
Ziyi Shen, Wei-Sheng Lai, Tingfa Xu, Jan Kautz, and Ming-Hsuan Yang.
\newblock Deep semantic face deblurring.
\newblock In {\em CVPR}, pages 8260--8269, 2018.

\bibitem{shocher2018zero}
Assaf Shocher, Nadav Cohen, and Michal Irani.
\newblock ``zero-shot'' super-resolution using deep internal learning.
\newblock In {\em ICCV}, pages 3118--3126, 2018.

\bibitem{singh2014super}
Abhishek Singh, Fatih Porikli, and Narendra Ahuja.
\newblock Super-resolving noisy images.
\newblock In {\em CVPR}, pages 2846--2853, 2014.

\bibitem{siu2012review}
Wan-Chi Siu and Kwok-Wai Hung.
\newblock Review of image interpolation and super-resolution.
\newblock In {\em The 2012 Asia Pacific Signal and Information Processing
  Association Annual Summit and Conference}, pages 1--10. IEEE, 2012.

\bibitem{sun2011learning}
Jian Sun and Marshall~F Tappen.
\newblock Learning non-local range markov random field for image restoration.
\newblock In {\em CVPR}, pages 2745--2752, 2011.

\bibitem{timofte2017ntire}
Radu Timofte, Eirikur Agustsson, Luc Van~Gool, Ming-Hsuan Yang, and Lei Zhang.
\newblock Ntire 2017 challenge on single image super-resolution: Methods and
  results.
\newblock In {\em CVPRW}, pages 114--125, 2017.

\bibitem{timofte2014a+}
Radu Timofte, Vincent De~Smet, and Luc Van~Gool.
\newblock A+: Adjusted anchored neighborhood regression for fast
  super-resolution.
\newblock In {\em ACCV}, pages 111--126, 2014.

\bibitem{venkatakrishnan2013plug}
Singanallur~V Venkatakrishnan, Charles~A Bouman, and Brendt Wohlberg.
\newblock Plug-and-play priors for model based reconstruction.
\newblock In {\em IEEE Global Conference on Signal and Information Processing},
  pages 945--948, 2013.

\bibitem{wang2018esrgan}
Xintao Wang, Ke Yu, Shixiang Wu, Jinjin Gu, Yihao Liu, Chao Dong, Yu Qiao, and
  Chen~Change Loy.
\newblock {ESRGAN}: Enhanced super-resolution generative adversarial networks.
\newblock In {\em The ECCV Workshops}, 2018.

\bibitem{yang2014single}
Chih-Yuan Yang, Chao Ma, and Ming-Hsuan Yang.
\newblock Single-image super-resolution: A benchmark.
\newblock In {\em ECCV}, pages 372--386, 2014.

\bibitem{yang2008image}
Jianchao Yang, John Wright, Thomas Huang, and Yi Ma.
\newblock Image super-resolution as sparse representation of raw image patches.
\newblock In {\em CVPR}, pages 1--8, 2008.

\bibitem{sun2016deep}
Yan Yang, Jian Sun, Huibin Li, and Zongben Xu.
\newblock Deep {ADMM-Net} for compressive sensing {MRI}.
\newblock In {\em NeurIPS}, pages 10--18, 2016.

\bibitem{yasarla2019deblurring}
Rajeev Yasarla, Federico Perazzi, and Vishal~M Patel.
\newblock Deblurring face images using uncertainty guided multi-stream semantic
  networks.
\newblock {\em arXiv:1907.13106}, 2019.

\bibitem{zhang2018ista}
Jian Zhang and Bernard Ghanem.
\newblock {ISTA-Net}: Interpretable optimization-inspired deep network for
  image compressive sensing.
\newblock In {\em CVPR}, pages 1828--1837, 2018.

\bibitem{zhang2015revisiting}
Kai Zhang, Xiaoyu Zhou, Hongzhi Zhang, and Wangmeng Zuo.
\newblock Revisiting single image super-resolution under internet environment:
  blur kernels and reconstruction algorithms.
\newblock In {\em PCM}, pages 677--687, 2015.

\bibitem{zhang2017learning}
Kai Zhang, Wangmeng Zuo, Shuhang Gu, and Lei Zhang.
\newblock Learning deep {CNN} denoiser prior for image restoration.
\newblock In {\em CVPR}, pages 3929--3938, July 2017.

\bibitem{zhang2018ffdnet}
Kai Zhang, Wangmeng Zuo, and Lei Zhang.
\newblock {FFDNet}: Toward a fast and flexible solution for {CNN}-based image
  denoising.
\newblock {\em IEEE TIP}, 27(9):4608--4622, 2018.

\bibitem{zhang2018learning}
Kai Zhang, Wangmeng Zuo, and Lei Zhang.
\newblock Learning a single convolutional super-resolution network for multiple
  degradations.
\newblock In {\em CVPR}, pages 3262--3271, 2018.

\bibitem{zhang2019deep}
Kai Zhang, Wangmeng Zuo, and Lei Zhang.
\newblock Deep plug-and-play super-resolution for arbitrary blur kernels.
\newblock In {\em CVPR}, pages 1671--1681, 2019.

\bibitem{zhang2019ranksrgan}
Wenlong Zhang, Yihao Liu, Chao Dong, and Yu Qiao.
\newblock Ranksrgan: Generative adversarial networks with ranker for image
  super-resolution.
\newblock In {\em ICCV}, pages 3096--3105, 2019.

\bibitem{zhang2018image}
Yulun Zhang, Kunpeng Li, Kai Li, Lichen Wang, Bineng Zhong, and Yun Fu.
\newblock Image super-resolution using very deep residual channel attention
  networks.
\newblock In {\em ECCV}, pages 286--301, 2018.

\bibitem{zhao2016fast}
Ningning Zhao, Qi Wei, Adrian Basarab, Nicolas Dobigeon, Denis Kouam{\'e}, and
  Jean-Yves Tourneret.
\newblock Fast single image super-resolution using a new analytical solution
  for $\ell2$-$\ell2$ problems.
\newblock {\em IEEE TIP}, 25(8):3683--3697, 2016.

\end{thebibliography}
}

\end{document}